\documentclass[aps,prx,reprint,showpacs,
superscriptaddress,longbibliography,nofootinbib] {revtex4-1}
\usepackage{graphicx}
\usepackage{dcolumn}
\usepackage{bm}

\usepackage{afterpage}
\usepackage{float}
\usepackage{placeins}
\usepackage{dsfont}
\usepackage[utf8]{inputenc}
\usepackage{bm}
\usepackage{amssymb}
\usepackage{amsmath}
\usepackage{amsfonts}
\usepackage{graphicx}
\usepackage[usenames,dvipsnames]{xcolor} 
\usepackage{dcolumn}
\usepackage{bm}

\usepackage[bookmarks=true,colorlinks,citecolor=blue,urlcolor=blue]{hyperref}

\usepackage{bbm}
\usepackage{float}

\usepackage{dcolumn}   
\usepackage{bm}        
\usepackage{filecontents}
\usepackage{lineno}

\usepackage{mathtools}
\usepackage[usenames,dvipsnames]{xcolor} 

\usepackage[export]{adjustbox}
\usepackage{adjustbox}

\usepackage{braket}
\usepackage{tikz}
\usepackage{enumitem}

\newcommand\bs[1]{\ensuremath{\boldsymbol{#1}}}

\newcommand{\eq}[1]{Eq.\thinspace(\ref{#1})}
\newcommand{\fig}[1]{Fig.\thinspace{}\ref{#1}}
\newcommand{\fc}[1]{({#1})}
\newcommand{\figc}[2]{Fig.\thinspace{}\ref{#1}\thinspace{}\fc{#2}}

\usetikzlibrary{arrows,shapes,positioning}
\usetikzlibrary{decorations.markings}
\tikzstyle arrowstyle=[scale=1]
\tikzstyle directed=[postaction={decorate,decoration={markings,
    mark=at position 0.8 with {\arrow[arrowstyle]{stealth}}}}]
\tikzstyle reverse directed=[postaction={decorate,decoration={markings,
    mark=at position 0.2 with {\arrowreversed[arrowstyle]{stealth};}}}]

\tikzstyle directeds=[postaction={decorate,decoration={markings,
    mark=at position 0.5 with {\arrow[arrowstyle]{stealth}}}}]
\tikzstyle reverse directeds=[postaction={decorate,decoration={markings,
    mark=at position 0.5 with {\arrowreversed[arrowstyle]{stealth};}}}]

\newcommand{\plaqv}{
\tikz[baseline=-0.62ex]{
\draw [gray] (0,-0.19) -- (0.3,-0.19);
\draw [gray] (0,0.19) -- (0.3,0.19);
\filldraw [color=black, fill=black!65] (0,0) ellipse (0.06 and 0.23); 
\filldraw [color=black, fill=black!65] (0.3,0) ellipse (0.06 and 0.23); }
}

\newcommand{\plaqh}{
\tikz[baseline=-0.62ex]{
\draw [gray] (-0.19,-0.19) -- (-0.19,0.19);
\draw [gray] (0.2,-0.19) -- (0.2,0.19);
\filldraw [color=black, fill=black!65] (0,-0.15) ellipse (0.23 and 0.058); 
\filldraw [color=black, fill=black!65] (0,0.15) ellipse (0.23 and 0.058); }
}

\newcommand{\loopP}{
\tikz[baseline=-0.62ex]{
\draw [gray] (0,-0.2) -- (0,0.2);
\draw [gray] (0.4,-0.2) -- (0.4,0.2);
\draw [black,ultra thick,directed] (0,-0.2) -- (0.4,-0.2);
\draw [black,ultra thick, reverse directed] (0,0.2) -- (0.4,0.2); }
}

\newcommand{\loopPP}{
\tikz[baseline=-0.62ex]{
\draw [gray] (0,-0.2) -- (0.4,-0.2);
\draw [gray] (0,0.2) -- (0.4,0.2);
\draw [black,ultra thick,directed] (0,-0.2) -- (0,0.2);
\draw [black,ultra thick, reverse directed] (0.4,-0.2) -- (0.4,0.2); }
}

\newcommand{\loopPPP}{
\tikz[baseline=-0.62ex]{
\draw [gray] (0,-0.2) -- (0,0.2);
\draw [gray] (0.4,-0.2) -- (0.4,0.2);
\draw [black,ultra thick,reverse directed] (0,-0.2) -- (0.4,-0.2);
\draw [black,ultra thick,directed] (0,0.2) -- (0.4,0.2); }
}

\newcommand{\loopPPPP}{
\tikz[baseline=-0.62ex]{
\draw [gray] (0,-0.2) -- (0.4,-0.2);
\draw [gray] (0,0.2) -- (0.4,0.2);
\draw [black,ultra thick,reverse directed] (0,-0.2) -- (0,0.2);
\draw [black,ultra thick,directed] (0.4,-0.2) -- (0.4,0.2); }
}

\definecolor{ilcharge}{RGB}{120,135,255}

\newcommand{\ilpair}{
\tikz[baseline=-0.62ex]{
\draw [gray] (0,0) -- (0.4,0); 
\node[mark size=2pt,color=ilcharge] at (0,0) {\pgfuseplotmark{*}};
\node[mark size=2pt,color=ilcharge] at (0.4,0) {\pgfuseplotmark{*}};
\draw[thick] (0,0) circle (2pt);
\draw[thick] (0.4,0) circle (2pt); }
}

\newcommand{\ltrivial}{
\tikz[baseline=-0.62ex]{
\draw [black,ultra thick,reverse directeds] (0,0) -- (0.6,0);
\draw [black,ultra thick,directeds] (0,0) -- (0.6,0); }
}

\begin{document}

\title{Emergent Fracton Dynamics in a Non-Planar Dimer Model}
\author{Johannes Feldmeier}
\email[]{johannes.feldmeier@tum.de}
\author{Frank Pollmann}
\author{Michael Knap}
\affiliation{Department of Physics and Institute for Advanced Study, Technical University of Munich, 85748 Garching, Germany}
\affiliation{Munich Center for Quantum Science and Technology (MCQST), Schellingstr. 4, D-80799 M{\"u}nchen, Germany}
\date{\today}

\begin{abstract}
We study the late time relaxation dynamics of a pure $U(1)$ lattice gauge theory in the form of a dimer model on a bilayer geometry. To this end, we first develop a proper notion of hydrodynamic transport in such a system by constructing a global conservation law that can be attributed to the presence of topological solitons. The correlation functions of local objects charged under this conservation law can then be used to study the universal properties of the dynamics at late times, applicable to both quantum and classical systems. Performing the time evolution via classically simulable automata circuits unveils a rich phenomenology of the system’s non-equilibrium properties: For a large class of relevant initial states, local charges are effectively restricted to move along one-dimensional `tubes' within the quasi-two-dimensional system, displaying fracton-like mobility constraints. The time scale on which these tubes are stable diverges with increasing systems size, yielding a novel mechanism for non-ergodic behavior in the thermodynamic limit. We further explore the role of geometry by studying the system in a quasi-one-dimensional limit, where the Hilbert space is strongly fragmented due to the emergence of an extensive number of conserved quantities. This provides an instance of a recently introduced concept of `statistically localized integrals of motion', whose universal anomalous hydrodynamics we determine by a mapping to a problem of classical tracer diffusion. We conclude by discussing how our approach might generalize to study transport in other lattice gauge theories.
\end{abstract}

\maketitle

{
  \hypersetup{hidelinks}
  \tableofcontents
}

\section{Introduction} \label{sec:1}
In recent years, efforts to understand the dynamics of constrained many-body systems have unveiled a rich phenomenology of exotic nonequilibrium properties. While interacting systems are generally expected to thermalize acccording to the eigenstate thermalization hypothesis (ETH)~\cite{Dalessio2016_eth,deutsch1991_quantum,srednicki1994_chaos,rigol2008_thermalization,
kim2015_eth}, constrained models can escape this generic scenario, either avoiding thermalization altogether, or approaching thermal equilibrium in an anomalously slow fashion.
Recent examples include quantum many-body scars in systems of Rydberg atoms~\cite{bernien2017_51atom,turner2018_scars,choi2019_scars,ho2019_scars,turner2018_qscars,
ok2019_topo,schecter2019_scars}, slow dynamics in kinetically constrained models~\cite{vanHorssen2015_gmbl,lan2018_glassy,feldmeier2019_glassy,
pancotti2020_east,guardadosanchez2020_quench,lee2020_frustration}, or localization in fracton systems~\cite{Sala2020_ergodicity,khemani2020_shatter,scherg2020_kinetic} that are characterized by excitations with restricted mobility~\cite{chamon2005_glass,haah2011_code,yoshida2013_fractal,vijay2015_topo,prem2017_glassy,
nandkishore2019_fractons,pretko2020_fracton}. Similarly, systems featuring exotic multipole conservation laws~\cite{pretko2017_subdim,pretko2018_gaugprinciple,pretko_witten,williamson2019_fractonic} have recently been found to exhibit anomalously slow emergent hydrodynamics~\cite{Guardado20,gromov2020_fractonhydro,feldmeier2020anomalous,zhang_2020}.

One particularly important class of such constrained models are lattice gauge theories, where a local Gauss law constrains the system dynamics. In general, understanding the effects of such gauge constraints on nonequilibrium properties is a challenging task. Recent efforts in this context have e.g. pointed out the possibility of strict localization in coupled gauge-matter and pure gauge theories~\cite{smith2017_local,smith2018_dynloc,brenes2018_gauge,karpov2020_disorderfree}, akin to many-body localization (MBL)~\cite{altshuler2006_mbl,nandkishore2015_mbl,altman2015_mbl,schreiber2015_mbl}. As an immediate related question, we can ask whether the presence of local gauge constraints can also have a qualitative effect on the relaxation towards equilibrium even if the constraints are not sufficiently strong to localize the system. In particular, pure gauge theories with a \textit{static} electric charge background, which can often be mapped to equivalent loop or dimer models, lack an obvious choice of suitable observables to probe the late-time relaxation dynamics due to the absence of charge transport.

In this work, we study a particular $U(1)$ gauge theory, a bilayer dimer model, where this limitation can be circumvented due to the presence of topological soliton configurations formed by the gauge fields. These solitons correspond to so-called `Hopfions' that exist more generally in the cubic dimer model~\cite{freedman_hopfions,bednik_hopfions,bednik2019_neural}. We show that the associated soliton conservation assumes the form of a usual $U(1)$ conservation law in the bilayer geometry, and thus provides a way to define a notion of transport via suitably chosen local correlation functions. 
Due to its universality, the late-time emergent hydrodynamic relaxation can be studied qualitatively using numerically feasible classically simulable circuits, as has recently been applied in other constrained systems~\cite{Iaconis19,morningstar2020_kinetic,feldmeier2020anomalous,iaconis2020multipole}. 
Many of the results described below can thus alternatively be viewed through the lense of lattice gases or cellular automata, but extend to the late time behavior of quantum systems as well.

After introducing the bilayer dimer model and deriving the abovementioned global $U(1)$ conservation law in Sec.~\ref{sec:2}, we divide the analysis of its associated dynamics into two parts:
In the first part, Sec.~\ref{sec:fractons}, we consider the model with full quasi-two-dimensional extension and study the dynamics of initial states hosting a finite density of conserved fluxes. Most strikingly, the local charges associated to the global soliton conservation law display fracton-like dynamics: While they are immobile as single particle objects, composites of these charges are effectively confined to diffuse along one-dimensional tubes within the quasi-2D system. We provide an explanation of these results in terms of a large class of conserved quantities that exist in the system. Notably however, the confinement to such effective 1D tubes is not due to subsystem symmetry constraints. Rather, the timescale necessary for charges to escape the 1D tubes diverges with increasing system size, providing an intriguing instance of non-ergodic behavior induced by the local gauge constraints.
In the second part, Sec.~\ref{sec:q1D}, we then go on to study the dynamics of the model in a quasi-one-dimensional limit on an open-ended cylinder. In this case, the Hilbert space is fragmented into an exponential (in system size) number of disjoint subspaces and hosts statistically localized integrals of motion (SLIOMs) that where introduced recently for constrained systems~\cite{rakovszky2020_sliom}. We determine the generic hydrodynamics exhibited by such SLIOMs and find subdiffusive decay of local correlations that can be understood analytically through the mapping to a classical tracer diffusion problem.
Having analyzed the dynamics of the bilayer dimer model, we conclude with Sec.~\ref{sec:solitons} by demonstrating explicitly that the global $U(1)$ charge is equivalent to a topological soliton conservation law in the form of Hopfions.

\section{Model and Conservation Laws} \label{sec:2}

\begin{center}
\textit{Hamiltonian}
\end{center}

\begin{figure}[t]
\begin{center}
\includegraphics[trim={0cm 0cm 0cm 0cm},clip,width=.9\linewidth]{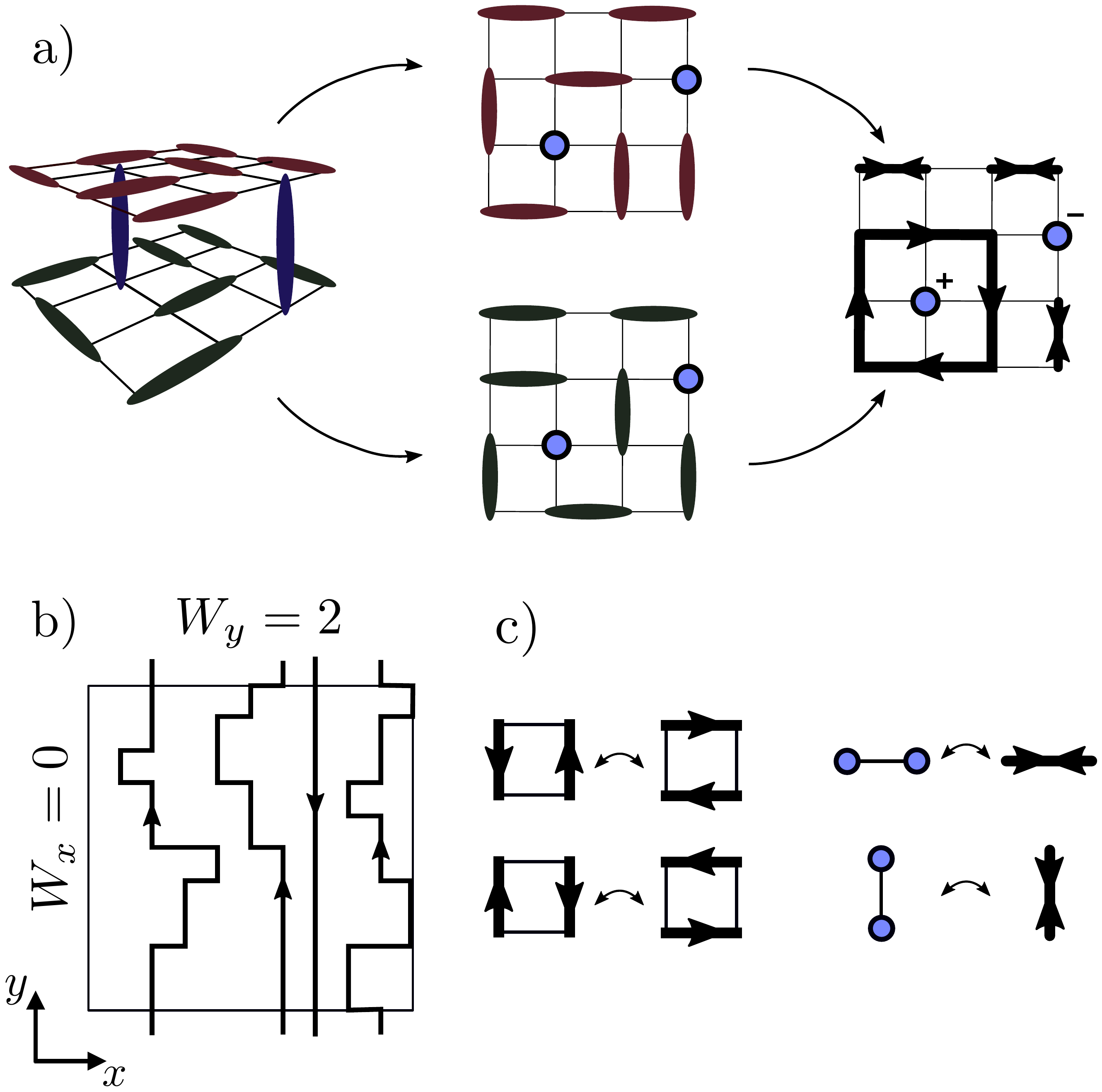}
\caption{\textbf{Construction of transition graphs.} \textbf{a)} The dimer configuration on the bilayer geometry is seperated into upper and lower layer, with interlayer dimers connecting the two marked as dots. The two configurations are then projected on top of each other from a bird's eye view to give rise to a directed loop model as explained in the main text. The interlayer dimers are assigned charges corresponding to their sublattice. The shown configuration is also the simplest one hosting a single Hopfion. \textbf{b)} Configurations with non-trivial fluxes $W_x$ and $W_y$ contain loops winding around the boundaries (other loops/interlayer charges not shown to avoid cluttering). \textbf{c)} Loop-moves originating from the elementary dimer plaquette flips.}
 \label{fig:1}
\end{center}
\end{figure}

The bilayer dimer model we consider is depicted in \figc{fig:1}{a}: It is given by two coupled layers of a square lattice, with bosonic hard-core dimers residing on the bonds, subject to a close-packing conditions of exactly one dimer touching each lattice site. If $\hat{n}^{(d)}_{\bs{r},\alpha} \in \{0,1\}$ denotes the dimer occupation on a bond $(\bs{r},\alpha)$ with $\alpha \in \{\pm x,\pm y,\pm z\}$, this condition can be phrased as 
\begin{equation} \label{eq:1}
\sum_{\alpha \in \{\mathrm{vertex} \, \bs{r}\}} \, \hat{n}^{(d)}_{\bs{r},\alpha} = 1 \quad \text{for all $\bs{r}$},
\end{equation}
where the sum extends over five nearest neighbor sites in \eq{eq:1} on the bilayer lattice. The Hilbert space of the system is then given by the set of all configurations satisfying \eq{eq:1} at every site.
The close-packing condition \eq{eq:1} can also be interpreted as a local gauge constraint, which explicitly turns into a Gauss law in a dual formulation of the quantum dimer model as an instance of a $U(1)$ quantum link model~\cite{chandrasekharan1997_link,wiese2013_link}. Details on the associated mapping for the planar case, which can easily be generalized to the (hyper)cubic geometry, can be found e.g. in Ref.~\cite{Celi2020_2Dgauge}.

With the Hilbert space at hand, we consider the standard Rokhsar-Kivelson (RK) model of elementary plaquette resonances between pairs of parallel dimers, which takes the pictorial form
\begin{equation} \label{eq:2}
\hat{H}_J = -J \sum_{p} \hat{h}_p = -J\sum_{p} \left(\ket{\plaqv}\bra{\plaqh} + h.c. \right).
\end{equation}
Here, the sum extends over all elementary plaquettes $p$ of the bilayer lattice. We can further allow for a constant potential term yielding an energy offset for each parallel dimer pair,
\begin{equation} \label{eq:3}
\hat{H}_{V} = V \sum_{p} \bigl( \hat{h}_p \bigr)^2 = V \sum_{p} \left(\ket{\plaqh}\bra{\plaqh} + \ket{\plaqv}\bra{\plaqv}\right),
\end{equation}
such that the full Hamiltonian is given by $\hat{H}=\hat{H}_J+\hat{H}_V$.

\begin{center}
\textit{Transition graph mapping and flux sectors}
\end{center}

We want to analyze the structure of the Hilbert space under the dynamics of \eq{eq:2} and to this end introduce a description in terms of an effective loop model. Such a description is known as `transition graphs', which we generalize here to the bilayer case. In this mapping, we take the two dimer configurations on the upper and lower layer and form their transition graph by projecting them on top of each other, see \figc{fig:1}{a}. This yields a model of closed, non-intersecting loops in the resulting projected two-dimensional layer. The smallest possible loop of length two consists of two horizontal dimers in both layers directly on top of each other. Notice that the loops can be assigned a chirality, which is inverted upon exchanging the configurations of upper and lower layer.

The directed loop segments can be described formally by new occupation numbers $\hat{n}^{(l)}_{\bs{r},\alpha} \in \{0,1\}$, which indicate the presence of a loop segment pointing from site $\bs{r}$ to $\bs{r}+\bs{e}_{\alpha}$, where $\bs{r}$ is now a \textit{two-dimensional} vector and $\alpha \in \{\pm x, \pm y\}$.
A full loop $\mathcal{L}$ of length $|\mathcal{L}|$ is then characterized by an ordered set $\bigl\{\bs{r}_0,\bs{r}_1,...,\bs{r}_{|\mathcal{L}|-1} \bigr\}$ of lattice sites, with $\bs{r}_{n+1}=\bs{r_n}+\bs{e}_\alpha$.
By convention, we choose the direction of a loop running through a site $\bs{r}=(r_x,r_y)$ as the orientation of the original dimer that occupies the site $\bs{r}^\prime=(\bs{r},r_z)$ with $r_z$ such that $\bs{r}^\prime$ is on the even (or A) sublattice, see \figc{fig:1}{a}.
The dimers between the two layers now appear as on-site particles for which we define the corresponding number operators $\hat{n}^{(h)}_{\bs{r}}$. We then assign a \textit{charge} to these particles depending on the sublattice they occupy, i.e. particles on sublattice $A$($B$) carry a charge $+1$($-1$). The total charge in the system is then given by $\sum_{\bs{r}} (-1)^{r_x+r_y}\, \hat{n}^{(h)}_{\bs{r}} = 0$, and we refer to the $\hat{n}^{(h)}_{\bs{r}}$ as `interlayer charges' in the following. A simple example of this construction is displayed in \figc{fig:1}{a}. 

Importantly, on periodic boundary conditions, there can exist non-local loops winding around the system boundaries, see \figc{fig:1}{b}. We can thus define global \textit{winding numbers} or \textit{fluxes} (these terms will be used interchangeably in this work) $W_x$ and $W_y$ for a given configuration by summing up the windings of all individual loops along both the $x$- and $y$-direction, respectively, see \figc{fig:1}{b}. The fluxes $W_x$ and $W_y$ are independently conserved under the dynamics of $\hat{H}_J$ and divide the Hilbert space into disconnected subspaces. In later sections, we will mainly be concerned with the additional structure of the Hilbert space on top of these flux sectors.

Finally, we can also translate the elementary plaquette flips of \eq{eq:2} to the loop picture, which take the form
$\hat{H}_J = \hat{H}_J^{(l)} + \hat{H}_J^{(h)}$,
where
\begin{equation} \label{eq:14}
\begin{split}
\hat{H}_J^{(l)} = \sum_p \Bigl[ \ket{\loopP} \bra{\loopPP} + \ket{\loopPPP} \bra{\loopPPPP} + h.c. \Bigr],
\end{split}
\end{equation}
and
\begin{equation} \label{eq:15}
\begin{split}
\hat{H}_J^{(h)} = \sum_{\left< \bs{r},\bs{r}^\prime \right>} \Bigl[ \ket{\ilpair} \bra{\ltrivial} + h.c. \Bigr].
\end{split}
\end{equation}
$\hat{H}_J^{(l)}$ describes the dynamics involving loop segments only, while $\hat{H}_J^{(h)}$ corresponds to the creation/annihilation of a $\pm$ interlayer charge pair on neighbouring sites, annihilating/creating a length-two loop on the same sites. In the pictorial representation of \eq{eq:15}, the charges from interlayer dimers are marked as blue circles in the transition graph.

\begin{center}
\textit{A global $U(1)$ conservation law}
\end{center}

\begin{figure}[t]
\begin{center}
\includegraphics[trim={0cm 0cm 0cm 0cm},clip,width=.99\linewidth]{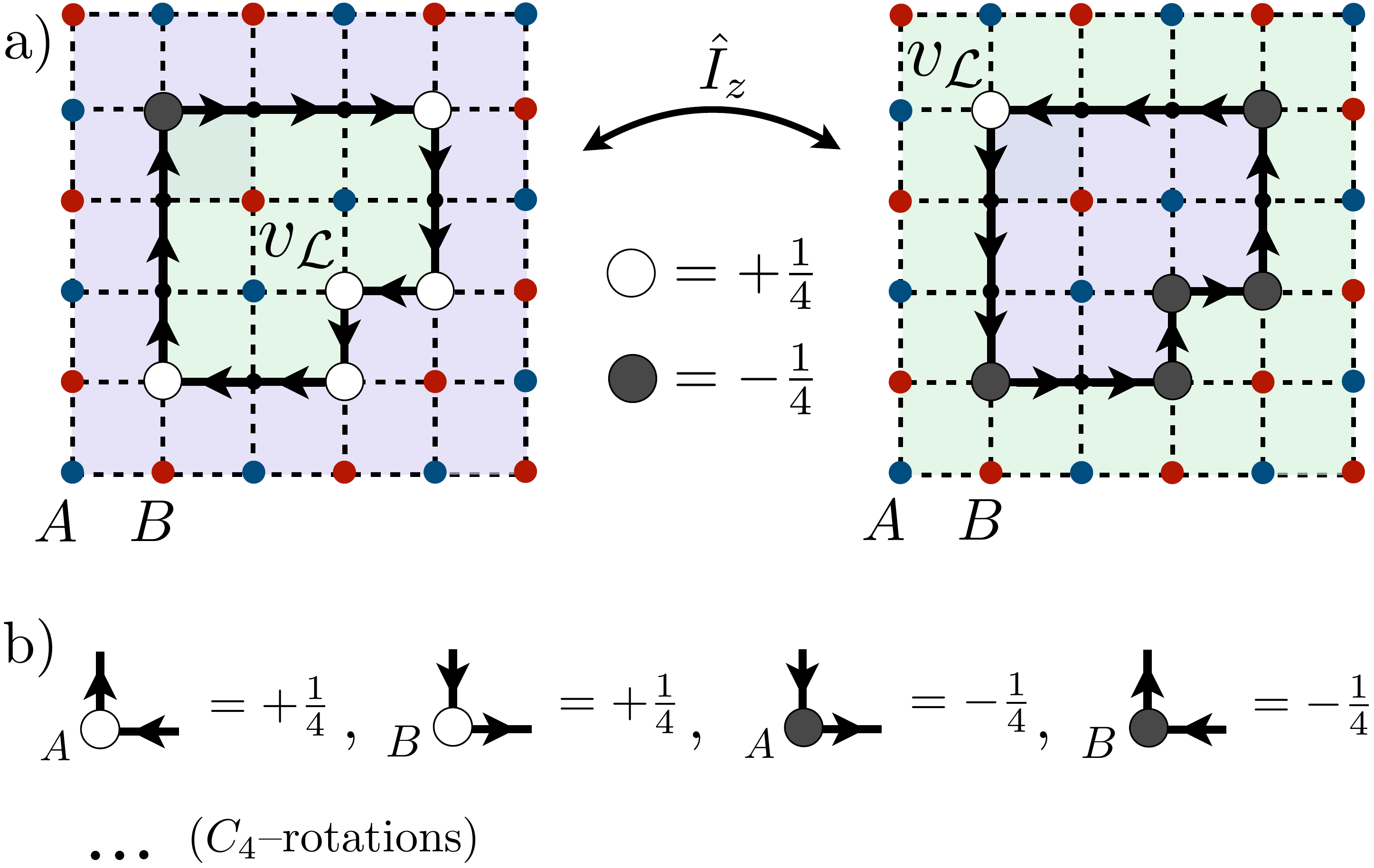}
\caption{\textbf{Loop interior and corner charges.} \textbf{a)} The interior $v_\mathcal{L}$ of a loop $\mathcal{L}$ contains all sites within the green shaded area, and turns into the complement upon inverting the chirality. In transition graphs, the loop chirality can be inverted via the inversion operator $\hat{I}_z$ which exchanges top and bottom layer of the original dimer lattice. White and gray circles illustrate the corner charges $\hat{\mathcal{C}}_{\bs{r}}=\pm \frac{1}{4}$ from \eq{eq:S3.5}. The validity of \eq{eq:S3.4} can directly be verified for this example. \textbf{b)} Dictionnary for the corner charges $\hat{C}_{\bs{r}}$, modulo lattice $C_4$--rotations.}
 \label{fig:3}
\end{center}
\end{figure}

The transition graph picture provides an intuitive starting point for deriving a global conserved charge $\hat{\mathcal{Q}}$ that we later, in Sec.~\ref{sec:solitons}, associate with the presence of topological solitons. Here, we first notice that under the loop dynamics of \eq{eq:14}, the difference in the number of $A$-- and $B$--sublattice sites enclosed by a particular loop stays constant as long as that loop does not split or merge with another loop. If such a split or merger occurs, then \textit{the sum} of the differences of $A$ and $B$ sites enclosed by the involved loops stays constant. Thus, if $v_{\mathcal{L}} \subset \mathbb{Z}^2$ denotes the interior of a loop $\mathcal{L}$, see \fig{fig:3}, then we can infer the global conserved quantity
\begin{equation} \label{eq:S3.2}
\hat{\mathcal{Q}} = \sum_{\mathcal{L}} \Delta N^{}_{AB}(v_\mathcal{L}),
\end{equation}
where the sum extends over all loops $\mathcal{L}$ in the transition graph of a given dimer configuration and $\Delta N^{}_{AB}(v_\mathcal{L})$ is the difference between the number of $A$ and $B$ sites contained in the set $v_{\mathcal{L}}$. We note that due to the Gauss law \eq{eq:1}, each site in the transition graph is either part of a loop or occupied by an interlayer charge. Since loops always contain an equal number of $A$ and $B$ sites, $\Delta N^{}_{AB}(v_\mathcal{L})$ is just the total interlayer charge enclosed by $\mathcal{L}$.

Since the loops $\mathcal{L}$ can become arbitrarily extended, \eq{eq:S3.2} is not in the form of a sum over local terms. However, for \textit{any} directed, closed, and non-intersecting loop $\mathcal{L}=\{\bs{r}_0,\bs{r}_1,...,\bs{r}_{|\mathcal{L}|-1}\}$ on the square lattice, the difference in the number of $A/B$ sites within $v_\mathcal{L}$ can be expressed as (see Appendix~\ref{sec:app3} for a proof)
\begin{equation} \label{eq:S3.4}
N^{}_{AB}(v_{\mathcal{L}}) = \frac{1}{4} \sum_{n=0}^{|\mathcal{L}|-1} (-1)^{x_n+y_n}\, \Bigl( \bs{\ell}^{}_{o}(\bs{r}_n) \wedge \bs{\ell}^{}_{i}(\bs{r}_n) \Bigr),
\end{equation}
where $\bs{\ell}^{}_{o}(\bs{r}_n)=\bs{r}_{n+1}-\bs{r}_n$ and $\bs{\ell}^{}_{i}(\bs{r}_n)=\bs{r}_n-\bs{r}_{n-1}$ are the directions of the out- and ingoing loop segments at $\bs{r}_n$. The symbol `$\wedge$' denotes the wedge product $\bs{a}\wedge \bs{b}=a_xb_y-a_yb_x$. \eq{eq:S3.4} is illustrated in \figc{fig:3}{a} with a specific example.
Using \eq{eq:S3.4} in \eq{eq:S3.2}, the quantity $\hat{\mathcal{Q}}$ can finally be expressed as
\begin{equation} \label{eq:S3.5}
\hat{\mathcal{Q}} =\frac{1}{4} \sum_{\bs{r}} (-1)^{r_x+r_y} \Bigl(  \hat{\bs{\ell}}_o(\bs{r}) \wedge \hat{\bs{\ell}}_i(\bs{r}) \Bigr) =: \sum_{\bs{r}} \hat{\mathcal{C}}_{\bs{r}},
\end{equation}
with the vector-valued operators 
\begin{equation}
\begin{split}
\hat{\bs{\ell}}_o(\bs{r}) &= \sum_{\alpha \in \{\pm x, \pm y\}}\bs{e}_\alpha \; \hat{n}^{(l)}_{\bs{r},\alpha} \\
\hat{\bs{\ell}}_i(\bs{r}) &= \sum_{\alpha \in \{\pm x, \pm y\}}\bs{e}_\alpha \; \hat{n}^{(l)}_{\bs{r}-\bs{e}_\alpha,\alpha}.
\end{split}
\end{equation}
\eq{eq:S3.5} assumes a particularly useful form, as $\hat{\mathcal{Q}}$ is now a sum over local terms $\hat{\mathcal{C}}_{\bs{r}}$. Since $\hat{\mathcal{C}}_{\bs{r}}\neq 0$ only when there is a corner of some loop at $\bs{r}$, we refer to the $\hat{\mathcal{C}}_{\bs{r}}$ as \textit{corner charges}, see \figc{fig:3}{b} for the specific relation between loop corners and the corresponding charge values. A more direct proof of $[\hat{H}_J,\hat{\mathcal{Q}}]=0$, regardless of the boundary conditions, is provided in Appendix~ \ref{sec:app4}. Moreover, inspecting \figc{fig:3}{a}, we see that a local excess of corner charges is directly connected to a local excess of interlayer charges. Finally, we notice that the corner charges carry only a fractional charge of $\pm 1/4$ and cannot move as independent particles, thus featuring fracton-like mobility constraints.

\begin{center}
\textit{Conserved chiral subcharges}
\end{center}

\begin{figure}[b]
\begin{center}
\includegraphics[trim={0cm 0cm 0cm 0cm},clip,width=.99\linewidth]{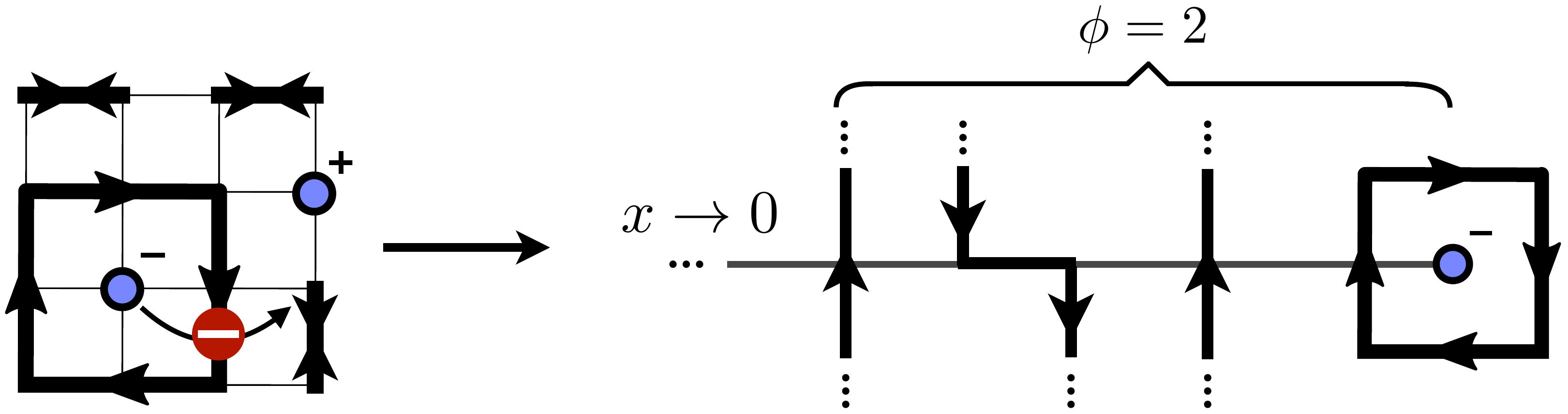}
\caption{\textbf{Construction of chiral subcharges.} An interlayer charge enclosed by a chiral loop cannot escape the loop under the dynamics of the Hamiltonian. We can then attach a string to an interlayer charge which extends all the way to the system boundary and determine the total chirality $\phi$ of all loops enclosing the charge. The sum of all interlayer charges with a fixed chirality $\phi$ then presents a conserved quantity.}
 \label{fig:2}
\end{center}
\end{figure}

As it turns out, there exists an even larger set of additional conserved quantities on top of the charge $\hat{\mathcal{Q}}$. To see this, let us recall that according to the previous considerations, an interlayer charge enclosed by a loop cannot exit the loop under the dynamics generated by Eqs.~(\ref{eq:14}),(\ref{eq:15}), see \fig{fig:2}.
To each interlayer charge, we can then associate a chirality via the total chirality of all loops enclosing it. Formally, we attach a string to the interlayer charge which extends all the way to the left system boundary and count the directed number of loop segments crossing it. For this purpose, we define the string operator
\begin{equation}  \label{eq:18}
\begin{split}
\hat{\phi}_{\bs{r}} &= \sum_{r^\prime_x = 0}^{r_x-1} \Bigl[ \hat{n}^{(l)}_{(r^\prime_x,r_y),y}
- \hat{n}^{(l)}_{(r^\prime_x,r_y+1),-y} \Bigr],
\end{split}
\end{equation}
that performs this counting, see \fig{fig:2}. We further define an associated chiral interlayer charge operator
\begin{equation}  \label{eq:19}
\begin{split}
\hat{q}_{\bs{r}}(\phi)= \hat{n}^{(h)}_{\bs{r}}\; \delta\bigl(\hat{\phi}_{\bs{r}} - \phi\bigr),
\end{split}
\end{equation}
that measures whether a given site $\bs{r}$ is occupied by an interlayer charge with chiral index $\phi$.


As a given interlayer charge cannot exit the loops enclosing it, it cannot interact with interlayer charges outside these loops. Thus, the only way to annihilate the interlayer charge is via the interaction with an opposite interlayer charge carrying \textit{the same} chiral index. Formally, in the notation introduced above, the following set of quantities are then \textit{independently} conserved under the Hamiltonian dynamics,
\begin{equation} \label{eq:20}
\hat{Q}_{\phi} = \sum_{\bs{r}} \, (-1)^{r_x+r_y}\, \hat{q}_{\bs{r}}(\phi)\; ; \qquad  \phi \, \in \, \Bigl\{-L_x,...,L_x\Bigr\}.
\end{equation}
We call the quantities $\hat{Q}_\phi$ `conserved chiral subcharges'. The formal proof of the invariance of \eq{eq:20}, by direct computation of the commutator $[\hat{H}_J,\hat{Q}_\phi]$, is given in Appendix~\ref{sec:app1}.
Importantly, the (non-local) chiral subcharges $\hat{Q}_\phi$ can also be related to the global quantity $\hat{\mathcal{Q}}$ via
\begin{equation} \label{eq:21}
\hat{\mathcal{Q}} = \sum_\phi \; \phi \; \hat{Q}_\phi,
\end{equation}
which we proof in Appendix~\ref{sec:app2}. The presence of these conserved subcharges will be crucial to understanding the resulting dynamics of corner charges both in Sec.~\ref{sec:fractons} and Sec.~\ref{sec:q1D}.

A few remarks are in order. While the above construction of the $\hat{Q}_\phi$ relied on open boundary conditions (at least in the $x$--direction), similar arguments proceed essentially analogously for periodic boundaries, where one can define relative instead of absolute chiralities of interlayer charges. We further note that the conservation of the quantities $\hat{Q}_\phi$ and $\hat{\mathcal{Q}}$ does depend on the dynamics being generated by elementary plaquette flips through \eq{eq:2} and, in general, does not persist in the presence of longer-range updates. However, such longer range updates are generally perturbatively small, and we may speculate that key features of the discussed physics still remain even in the presence of such terms.

\section{Emergent Fracton Dynamics in the 2D Bilayer Dimer Model} \label{sec:fractons}
\begin{figure}[t]
\begin{center}
\includegraphics[trim={0cm 0cm 0cm 0cm},clip,width=.99\linewidth]{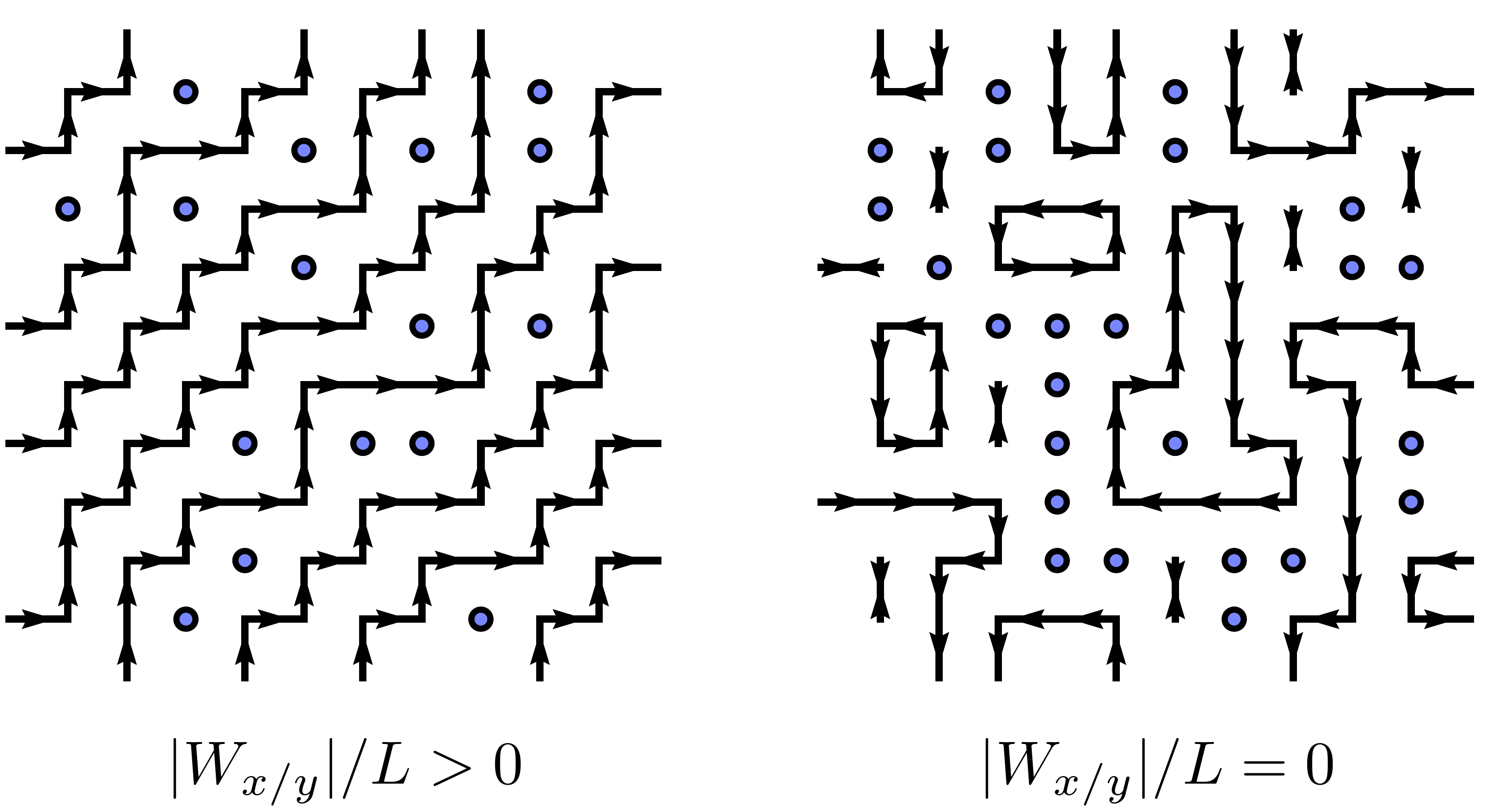}
\caption{\textbf{Initial states.} Local snapshots of typical example configurations within the transition graph picture, both for finite ($W_x/L_x=W_y/L_y>0$) and vanishing ($W_x/L_x=W_y/L_y=0$) flux densities.}
 \label{fig:S4.2}
\end{center}
\end{figure}

Having derived the conserved quantity $\hat{\mathcal{Q}}$ in the form of \eq{eq:S3.5}, we are interested in how the associated local corner charges $\hat{\mathcal{C}}_r$ are transported through the system under a generic time evolution.
Notice that any nonequilibrium dynamics within the dimer Hilbert space, either from $e^{-i\hat{H}t}$ or some other unitary evolution built up by elementary plaquette flips, depends in general on the chosen initial state. In the following, we will focus on the real time dynamics emerging from initial states that host a finite density $|W_x|/L=|W_y|/L > 0$ of fluxes, see \fig{fig:S4.2}. We remark that such initial states can also be generated thermodynamically at low energies of a classical dimer model with energy function $\hat{H}_V$, which we have verified using Monte Carlo simulations.


\subsection{Time evolution}

\begin{figure}[b]
\begin{center}
\includegraphics[trim={0cm 0cm 0cm 0cm},clip,width=.99\linewidth]{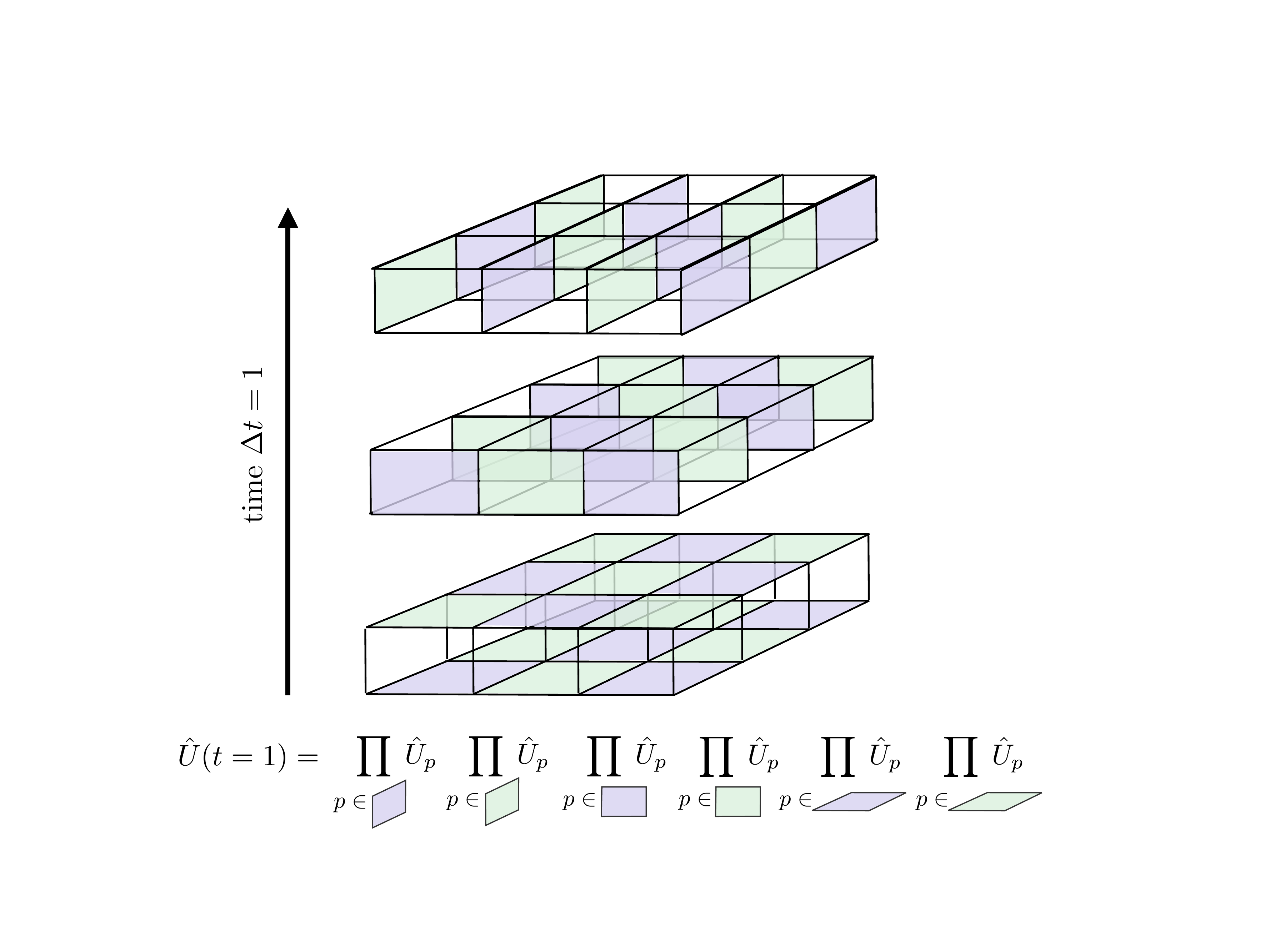}
\caption{\textbf{Time evolution.} A single time step of the deterministic unitary evolution, built on the elementary plaquette updates $\hat{U}_p$ of \eq{eq:S4.1}. Within a fixed plaquette color, all associated local updates commute.}
 \label{fig:S4.3}
\end{center}
\end{figure}

\begin{figure*}[t]
\begin{center}
\includegraphics[trim={0cm 0cm 0cm 0cm},clip,width=.9\linewidth]{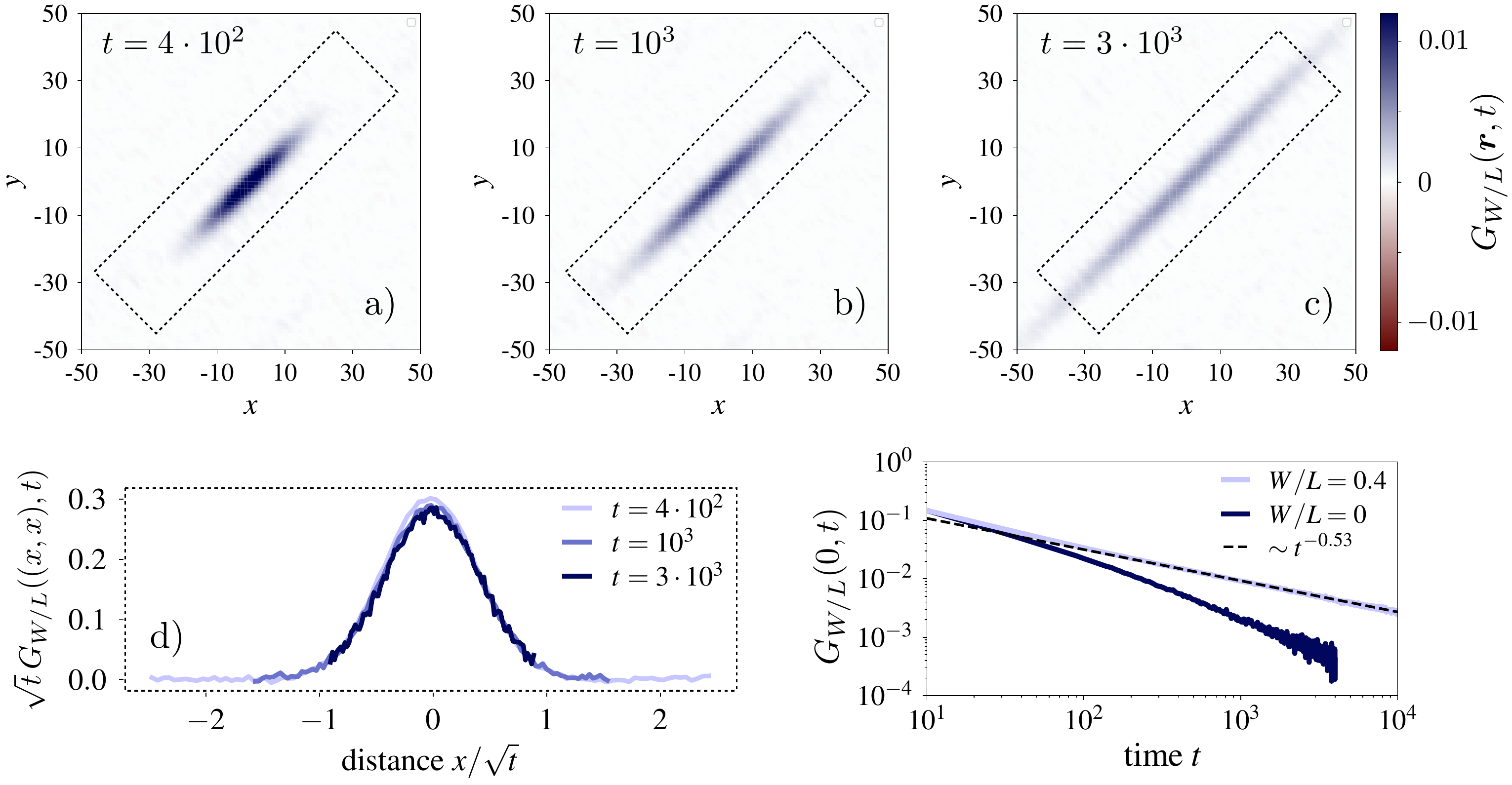}
\caption{\textbf{Relaxation dynamics of corner charge correlations.} \textbf{a)}+\textbf{b)}+\textbf{c)} The spatially resolved corner charge correlation function $G_{W/L}(\bs{r},t)=\braket{\hat{\mathcal{C}}_{0}(0)\,\hat{\mathcal{C}}_{\bs{r}}(t)}_{W/L}$ at different times $t$. The initial states in the average $\braket{...}_{W/L}$ are sampled from random states within the fixed winding sector $W_x/L=W_y/L=0.4$. We observe a clear restriction of the charge dynamics to an effective one-dimensional `tube' along the diagonal. \textbf{d)} The dynamics along the diagonal is diffusive, demonstrated by the scaling collapse to a one-dimensional (Gaussian) diffusion kernel. \textbf{e)} The return probability $\braket{\hat{\mathcal{C}}_{\bs{r}}(0)\,\hat{\mathcal{C}}_{\bs{r}}(t)}_{W/L}$ is anomalously slow for a two-dimensional system with $W/L=0.4$. In contrast, for a vanishing winding density $W/L=0$, the correlations decay fast. The system sizes are $L_x=L_y=200$ and $L_x=L_y=1000$ for $W/L=0.4$ and $W/L=0$, respectively.}
 \label{fig:4}
\end{center}
\end{figure*}

Let us now introduce the unitary time evolution that allows us to study the dynamics of corner charges. Ideally, one would like to consider the full quantum time evolution $e^{-i\hat{H}t}$ for the closed system dynamics. This, however, is a challenging task due to the large Hilbert space in our quasi-2D system. Instead, we use that for conserved quantities such as $\hat{\mathcal{Q}}$, an effectively classical hydrodynamic picture at late times is expected to emerge~\cite{chaikin_lubensky_1995,Mukerjee06,Lux14,Bohrdt16,leviatan2017quantum,Parker19,
Khemani2018,Rakovszky2018,Gopalakrishnan19,schuckert2020_hydro}. Due to this universal late time decay, every sufficiently ergodic time evolution that respects the system's conservation laws is expected to result in the same qualitative hydrodynamic tail. Details of the short time quantum coherent dynamics would therefore merely enter the numerical value of an effective diffusion constant. Thus, in order to capture only the qualitative aspects of the charge dynamics at late times, we can construct an alternative, classically simulable unitary evolution built up by elementary plaquette flips. This approach follows recent works on automata circuits~\cite{Iaconis19,iaconis2020quantum}, that have been applied to study the transport properties of fracton models~\cite{Iaconis19,morningstar2020_kinetic,feldmeier2020anomalous,iaconis2020multipole,
Guardado20,gromov2020_fractonhydro}, and have even been connected to the dynamics of more conventional random unitary quantum circuits~\cite{moudgalya2020_spectral}.

The elementary local unitary corresponding to a plaquette flip update is given by
\begin{equation} \label{eq:S4.1}
\hat{U}_p = \Bigl[ \mathds{1} - \bigl( \hat{h}_p \bigr)^2 \Bigr] + \hat{h}_p,
\end{equation}
with $\hat{h}_p$ from \eq{eq:2}. The action of \eq{eq:S4.1} on a given dimer configuration $\ket{\psi}$ (represented as a product state) is easily understood: If $\ket{\psi}$ has a flippable plaquette at $p$, then $\hat{U}_p \ket{\psi} = \hat{h}_p \ket{\psi}$, i.e. the plaquette is flipped and we obtain a new product state. If, however, $\ket{\psi}$ has no flippable plaquette at $p$, then $\hat{U}_p\ket{\psi} = \ket{\psi}$, i.e. the state remains unchanged. We can then use the elementary updates from \eq{eq:S4.1} as building blocks for defining a discrete time evolution scheme that can be simulated as a classical cellular automaton.
To this end, we can define a deterministic Floquet time evolution, where
\begin{equation} \label{eq:S4.3}
\hat{U}(t) = \left( \prod_{i=1}^{4L_xL_y} \hat{U}_{p_i} \right)^t,
\end{equation}
and the plaquettes $p_i$ are kept \textit{fixed} throughout different instances of the time evolution. Furthermore, the $p_i$ should be such that for $i\in \{1,...,4L_xL_y\}$, each plaquette appears exactly once within a Floquet period. We emphasize that alternative choices of update schemes, such as stochastic updates, yield a qualitatively equivalent late time relaxation, and throughout this work we employ a fixed deterministic evolution that is illustrated in \fig{fig:S4.3}.

Using these unitary evolution operators, we can then compute e.g. the correlation function of the previously introduced corner charges,
\begin{equation} \label{eq:S4.4}
G^{}_{S}(\bs{r},t) := \braket{\hat{\mathcal{C}}_{\bs{r}}(t)\, \hat{\mathcal{C}}_0(0)}_{S \subseteq \mathcal{H}},
\end{equation}
where the average $\braket{...}_{S \subseteq \mathcal{H}}$ is taken over dimer occupation number product initial states within some predefined subset $S \subseteq \mathcal{H}$ of the full Hilbert space.

\subsection{Reduced mobility of corner charges}

\begin{figure}[t]
\begin{center}
\includegraphics[trim={0cm 0cm 0cm 0cm},clip,width=.99\linewidth]{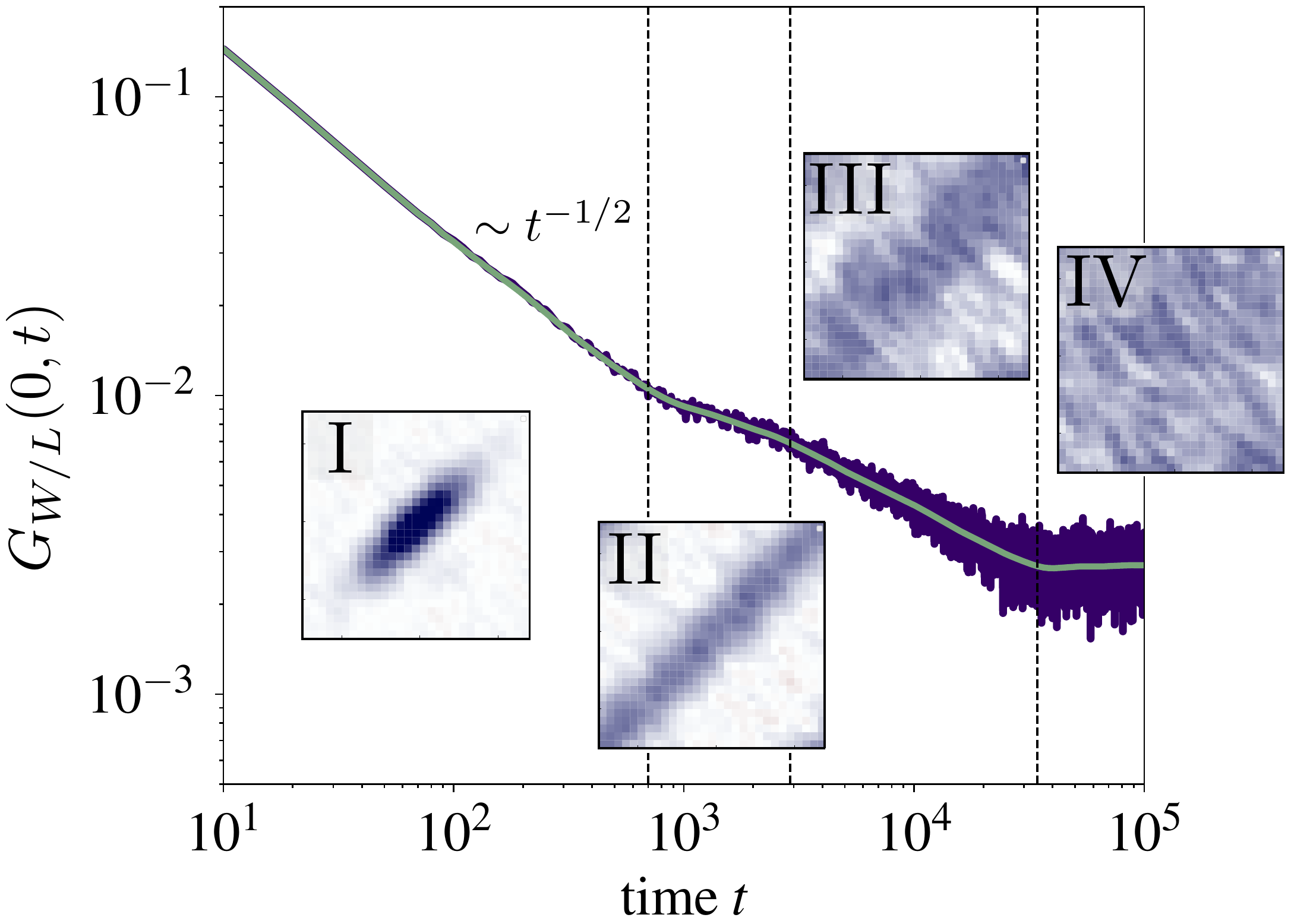}
\caption{\textbf{Thermalization process.} The way to equilibrium, i.e. full delocalization of corner charges across the 2D system, can be split into four distinct stages that are reflected in local correlation functions: I: A stage of diffusive dynamics along effective one-dimensional tubes, with $G\sim t^{-1/2}$. II: A plateau where the charge is fully delocalized along the tube. III: The delocalization along the second direction sets in. IV: The charge is fully delocalized across the whole system. The system size in this example is $L_x=L_y=30$ and the green curve corresponds to the moving average. This demonstrates that the formation of 1D tubes is not due to disconnectivities, but rather \textit{bottlenecks} in the Hilbert space structure.}
 \label{fig:5}
\end{center}
\end{figure}

\begin{figure}[t]
\begin{center}
\includegraphics[trim={0cm 0cm 0cm 0cm},clip,width=.99\linewidth]{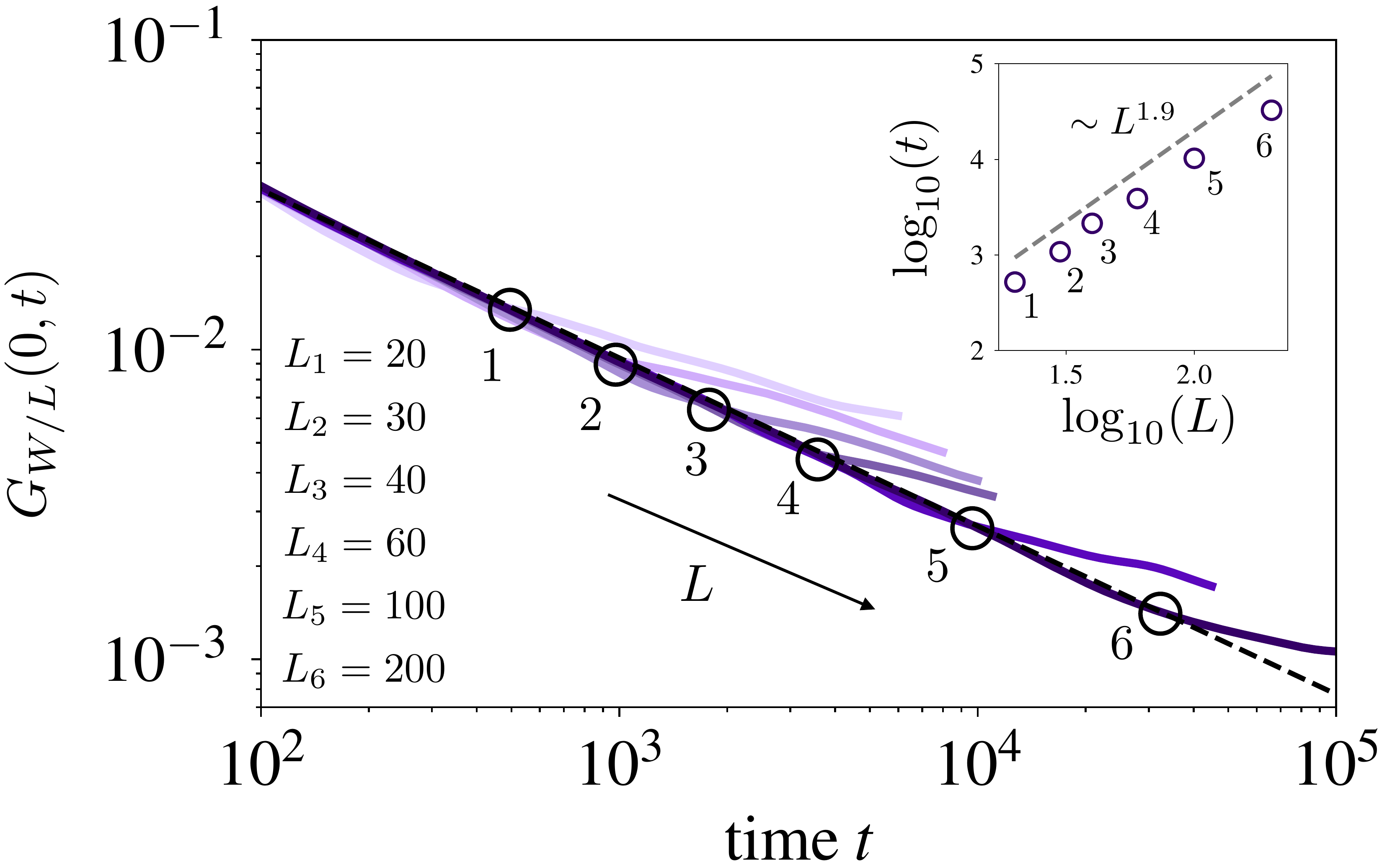}
\caption{\textbf{Finite size scaling.} For increasing system size, the correlations follow the diffusive decay for increasingly long times, approximately scaling as $\sim L^2$ as shown in the inset. Therefore, the one-dimensional tubes are expected to persist up to infinite time in the thermodynamic limit. The displayed lines correspond to the moving average of the numerically sampled correlations.}
 \label{fig:6}
\end{center}
\end{figure}

\begin{figure}[b]
\begin{center}
\includegraphics[trim={0cm 0cm 0cm 0cm},clip,width=.8\linewidth]{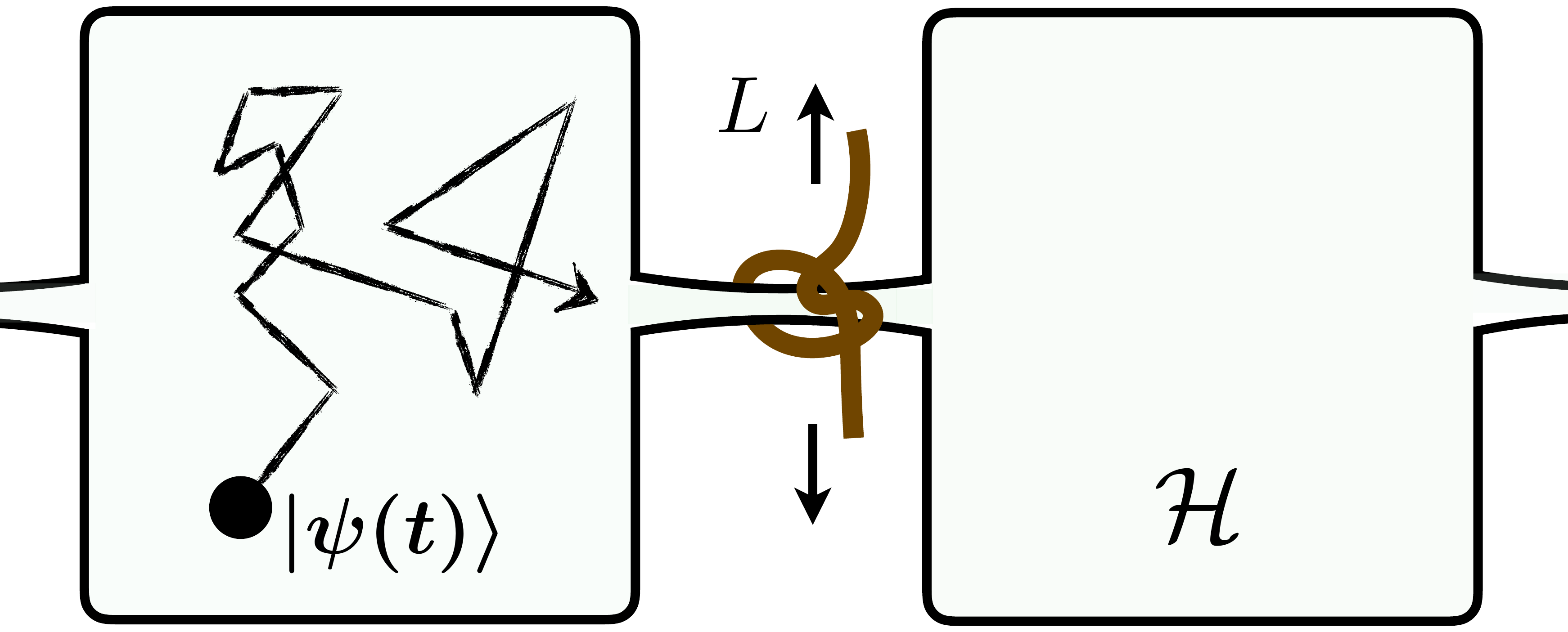}
\caption{\textbf{Origin of reduced dimensional mobility.} The numerical results presented in Sec.~\ref{sec:fractons} can be understood intuitively in terms of \textit{bottlenecks} in the Hilbert space structure: As the system size increases, the bottlenecks become narrower, and the time needed to eventually pass through the bottleneck in order to explore the full Hilbert space diverges.}
 \label{fig:S4.5}
\end{center}
\end{figure}

Having defined a proper time evolution, we move on to study to dynamics of corner charges via the correlations defined in \eq{eq:S4.4}. In particular, we will focus on averages over randomly chosen initial states hosting a finite flux density $W/L=W_x/L_x=W_y/L_y=0.4$. The associated correlations are then denoted as $G_{W/L}(\bs{r},t)$. The dynamics within such flux sectors is particularly interesting: As we saw in the construction of conserved quantities in Sec.~\ref{sec:2}, loops act as obstacles to the dynamics of both interlayer- and corner charges. The presence of non-contractible loops carrying a finite winding number should thus essentially trap charges in between two such winding loops. However, the loops themselves are dynamical objects as well, and we require the time evolution introduced above to resolve the ensuing system dynamics. We note that the resulting late time relaxation should then be qualitatively equivalent to the closed system quantum dynamics for $\hat{H}_V=0$, which, for product initial states at zero energy, corresponds to infinite temperature due to the symmetric spectrum of $\hat{H}_J$ (see Appendix~\ref{sec:app5}).

Our main numerical results for a system of size $L_x=L_y=200$ are presented in \fig{fig:4}. Inspecting the spatially resolved $G_{W/L}(\bs{r},t)$ in \figc{fig:4}{a-c}, we find diffusion of corner charges along effective, one-dimensional \textit{tubes} within the 2D system. The diagonal direction of these tubes within the system corresponds to the winding order of the initial states, cf. \fig{fig:S4.2}. In \figc{fig:4}{d}, we show the correlations $G_{W/L}(\bs{r}=(x,x),t)$ along the tube direction. These follow a 1D diffusion kernel $G_{W/L}(\bs{r}=(x,x),t) = \frac{1}{\sqrt{Dt}}e^{-x^2/Dt}$. From the viewpoint of the site-local return probability $G_{W/L}(0,t)\sim 1/\sqrt{t}$, this leads to anomalous slow decay of local correlation functions, which would generically be expected to relax as $\sim 1/t$ for usual diffusion in two dimensions. This is demonstrated in \figc{fig:4}{e}, where we compare $G_{W/L}(0,t)$ to the faster decaying correlations within the zero flux sector.

Remarkably, although the winding loops are dynamical as well and should in principle be able to move throughout the entire system, the effective 1D tubes do not appear to broaden within the times shown in the correlations of \figc{fig:4}{a-c}. Therefore, an important question concerns whether in the thermodynamic limit, there exists a finite (but potentially very large) timescale at which the localization of corner charges within stationary 1D tubes eventually breaks down. To this end, we consider the return probability $G_{W/L}(0,t)$ within a smaller system of size $L_x=L_y=30$ in \fig{fig:5}, which reveals a multistage thermalization process in this finite size system: First (see (I) in \fig{fig:5}), charges diffusive along the effective one-dimensional tubes. Then (see (II) in \fig{fig:5}), the system reaches an intermediate plateau where the charges are fully delocalized along the 1D tube, but still remain localized with respect to the perpendicular direction. Eventually (see (III) and (IV) in \fig{fig:5}), the 1D tubes start to broaden, and charges are delocalized across the entire 2D system. These results demonstrate that the winding loops are indeed in principle able to move through the system. To answer our question about the thermodynamic limit posed above, we then need to study how the different timescales involved in the multistage thermalization process of \fig{fig:5} change as we increase the system size.

This analysis is perfomed in \fig{fig:6}, where we show $G_{W/L}(0,t)$ for a range of system sizes $L \in \{20,30,40,60,100,200\}$. As we increase $L$, the return probability follows the 1D diffusive decay for increasingly long times. In particular, the largest system size $L=200$ still follows purely 1D relaxation at times when smaller systems have already fully delocalized. This suggests that in the thermodynamic limit, the time scale required to move the winding loops through the system indeed diverges. As a consequence, the system exclusively exhibits effectively 1D dynamics in the thermodynamic limit, failing to delocalize perpendicular to the winding direction.
Intuitively, the diverging timescale of eventual 2D delocalization can be understood by the fact that non-local winding loops have to be moved as a whole for such a process to occur. Since the length of these loops diverges with system size, the timescale of these processes diverges as well.

We emphasize that the reduced dimensionality found for the charge dynamics -- a hallmark of fracton--like excitations -- comes without the presence of subsystem symmetries that would fundamentally restrict the charges to only move along one dimension, as is evidenced by the eventual 2D decay in finite size systems. As the corner charges can in principle move through the entire system, the generator of the dynamics is not `reducible' in the language of classically constrained models~\cite{ritort2003_glassy}. Thus, instead of the Hilbert space falling into disconnected parts in the form of symmetry sectors, the fractonic behavior in the bilayer dimer model is rather due to bottlenecks in the Hilbert space, which become narrower as the system size is increased, see \fig{fig:S4.5} for a symbolic depiction of the situation. It would be interesting to see how such a Hilbert space structure effects the validity of the eigenstate thermalization hypothesis (ETH) with respect to the Hamiltonian $\hat{H}$.

\section{The Quasi 1D Bilayer Model} \label{sec:q1D}
In the previous section we have numerically demonstrated the emergence of reduced dimensional mobility for the corner charges of \eq{eq:S3.5} in translationally invariant 2D systems. In this section, we change the geometry and consider a \textit{quasi-one-dimensional}, cylindrical system. There, we encounter a strong fragmentation of the Hilbert space into an exponential in system size number of disconnected subsectors. In addition, the associated conserved quantities that label the different Hilbert space sectors fulfill a recently introduced concept of statistical localization~\cite{rakovszky2020_sliom}. We determine the algebraic long time decay of the corner charge correlations by mapping to a classical problem of tracer diffusion in a 1D system with hard core interacting particles.

\subsection{Hilbert space fragmentation for large flux}
We consider a quasi-1D system on a cylinder of length $L_x$ with open boundaries, whose circumference $L_y$ is kept finite. In order to analyze the structure of the Hilbert space within this geometry, we investigate the relative size of the different disconnected subspaces that each can be labelled by a set of values $\{Q^{}_\phi\}_{\phi \in [-L_x,L_x]}$ (with $Q^{}_\phi \in \mathbb{Z}$) of the conserved chiral subcharges of \eq{eq:20}. Implicitly assuming the flux number $W_y$ to be fixed, we denote the relative size of the $\{Q_\phi\}$--subspace compared to the full Hilbert space at flux $W_y$ by $P(\{Q_\phi\})$. We note that $P(\{Q_\phi\})$ is a probability distribution, i.e. $\sum_{\{Q_\phi\}}P(\{Q_\phi\}) = 1$, which can be sampled numerically by randomly drawing states from the full $W_y$--subspace. In addition, we can define the individual probabilities $P(Q_\phi) = \sum_{\{Q_{\phi^\prime}\}_{\phi^\prime \neq \phi}}\, P(\{Q_{\phi^\prime}\})$ for the $\phi$--th subcharge $\hat{Q}_\phi$ to assume a value $Q_\phi$.
It can be verified numerically through Monte Carlo sampling that different $\hat{Q}_\phi$ are essentially uncorrelated, i.e. $\braket{\hat{Q}_\phi \hat{Q}_{\phi^\prime}} \propto \delta_{\phi,\phi^\prime}$, which allows us to approximate $P(\{Q_\phi\}) \approx \prod_{\phi} P(Q_\phi)$ for the following arguments.

\begin{figure}[b]
\begin{center}
\includegraphics[trim={0cm 0cm 0cm 0cm},clip,width=.99\linewidth]{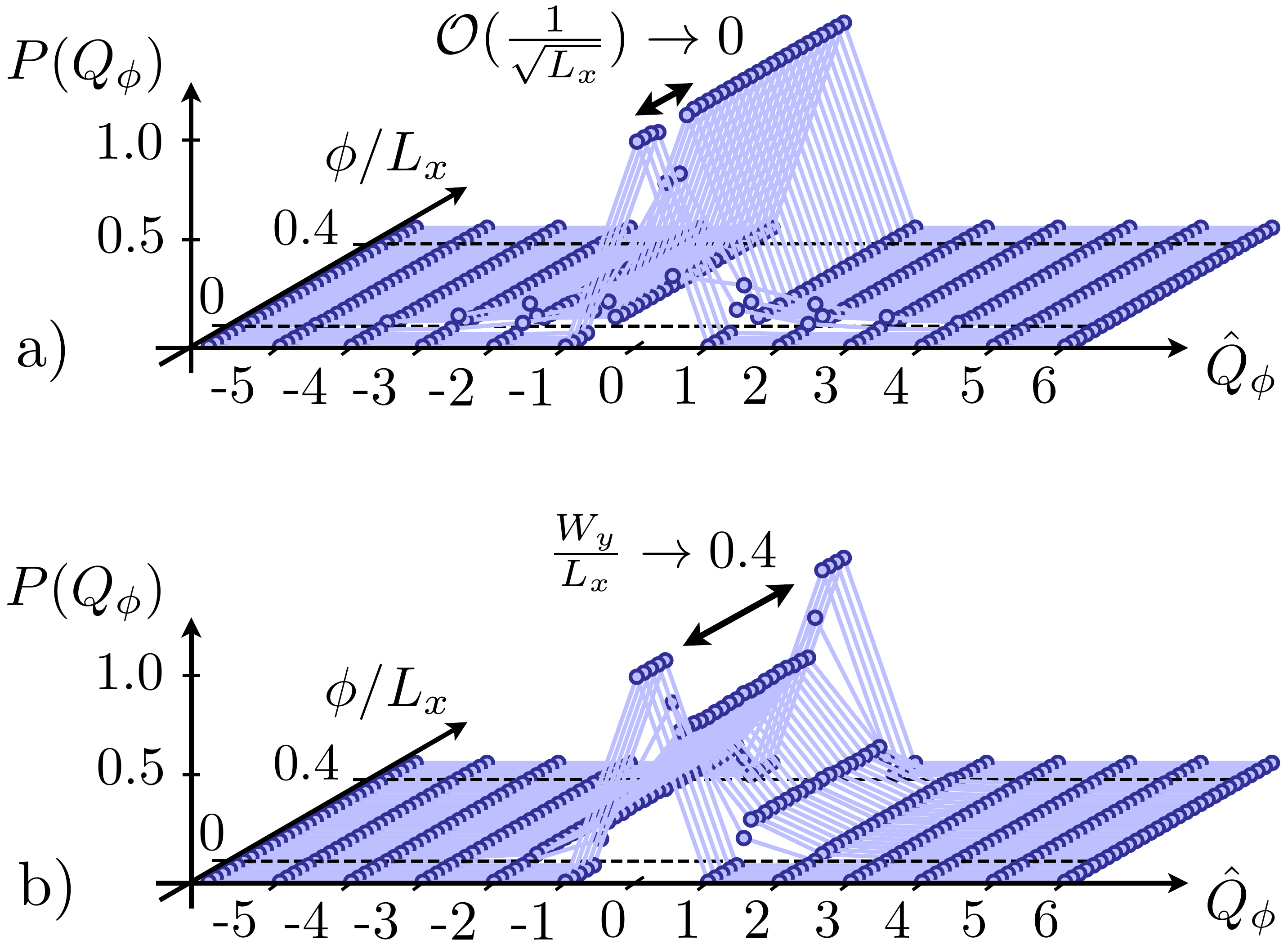}
\caption{\textbf{Hilbert space fragmentation in a quasi-1D geometry.} We display the probability distributions $P(\hat{Q}_\phi)$ of the charges $\hat{Q}_\phi$ for randomly chosen states on a $L_x=100$, $L_y=6$ open-ended cylinder. \textbf{a)} For vanishing flux $W_y=0$, almost all $\hat{Q}_\phi$ are statistically fixed to zero, i.e. $P(Q_\phi)=\delta^{}_{Q_\phi,0}$. Thus, a small number of large sectors dominates the Hilbert space, which is only weakly fragmented. \textbf{b)} In contrast, for a finite winding density $W_y/L_x=0.4$, there is an extensive amount ($0 \lesssim \phi \lesssim W_y$) of charges $\hat{Q}_\phi$ with a finite width probability distribution. As a result, the Hilbert space is strongly fragmented.}
 \label{fig:S5.1}
\end{center}
\end{figure}

Choosing the flux-free sector $W_y=0$ at first, we plot the distributions $P(Q_\phi)$ in \figc{fig:S5.1}{a}. We see that for almost all values of $\phi$, the values $Q_\phi$ of the subcharges are statistically fixed to zero, i.e. $P(Q_\phi)=\delta_{Q_\phi,0}$. This holds for all $\phi$ outside a range of order $\mathcal{O}(\sqrt{L_x})$ around $\phi=0$, which is expected from generic fluctuations of the distribution of winding loops even within the $W_y=0$~--~sector. Therefore, although an extensive number of conserved quantities $\hat{Q}_\phi$ exist, most of the associated subspaces are small, and the Hilbert space is instead dominated by a small number of very large sectors. In the terminology of Ref.~\cite{Sala2020_ergodicity}, the Hilbert space is only \textit{weakly fragmented}.

In contrast, for a finite flux density $W_y/L_x > 0$ around the cylinder, the probability distributions $P(Q_\phi)$ obtain a finite width for an extensive number of $\phi$ between $0 \lesssim \phi \lesssim W_y$, see \figc{fig:S5.1}{b}. As a consequence, the relative size of \textit{every} $\{Q_\phi\}$--subsector is exponentially suppressed with respect to the full Hilbert space: $P(\{Q_\phi\})\approx \prod_\phi P(Q_\phi) \sim e^{-c\, L_x}$, since an extensive number of the $P(Q_\phi)$ in the product over $\phi$ is smaller than one. In particular, also the largest sector, $\{Q_\phi=0\}$ for all $\phi$, is exponentially suppressed, which is seen intuitively by multiplying all the probabilities $P(Q_\phi=0)$ along the $\phi$--axis in \figc{fig:S5.1}{b}. According to the definition provided in Ref.~\cite{Sala2020_ergodicity} the Hilbert space is thus \textit{strongly fragmented}.

\subsection{Statistical localization of chiral subcharges}
Having identified the strong fragmentation of the Hilbert space in the previous section, we now turn to determine some of its consequences. In particular, the previous results imply the applicability of the recently introduced concept of statistical localization~\cite{rakovszky2020_sliom}. Let us shortly describe this concept in a hands-on way: The conserved chiral subcharges are given by $\hat{Q}_\phi = \sum_{\bs{r}}(-1)^{r_x+r_y}\,\hat{q}_{\bs{r}}(\phi)$, where $\hat{q}_{\bs{r}}(\phi)$ checks whether there is an interlayer charge at site $\bs{r}$ that is encircled by $\phi$ loops in total. Notice that within the cylindrical geometry, the definition of the $\hat{Q}_\phi$ remains well-defined. As there is a finite density of winding loops, we would then generally expect the main contributions to $\hat{Q}_\phi$ to come from interlayer charges located around the position of the $\phi^{th}$ winding loop. Where in turn is the $\phi^{th}$ winding loop located at along the cylinder? To get an estimate, let us assume there to be exactly $W_y$ winding loops and let us further ascribe a one-dimensional position $x_\phi$ to the $\phi^{th}$ such loop. Then the probability $p_\phi(x_\phi)$ of finding this loop at $x_\phi$ is approximated by a simple count of the number of possibilities: $p_\phi(x_\phi) \approx \frac{\binom{x_\phi}{\phi-1}\binom{L_x-x_\phi-1}{W_y-\phi-1}}{\binom{L_x}{W_y}}$, i.e. the number of possibilities to have $\phi-1$ winding loops to the left of $x_\phi$ times the number of possibilities to have the remaining $W_y-\phi-1$ loops to the right of $x_\phi$, divided by the overall number of possibilities to distribute the one-dimensional positions of all $W_y$ loops across the system of length $L_x$~\cite{rakovszky2020_sliom}. For a winding loop in the bulk of the system, $p(x_\phi)$ is a peaked distribution of width $\sqrt{L_x}$ centered around $\phi L_x/W_y$. Via the line of arguments just provided, we then expect a very similar distribution $p_\phi(x_\phi)$ to describe the locations of the operators $\hat{q}_{\bs{r}}(\phi)$. Therefore, the $\hat{q}_{\bs{r}}(\phi)$ that constitute the $\hat{Q}_\phi$ are localized to a \textit{subextensive} region of size $\sqrt{L_x}$, a feature termed statistical localization in Ref.~\cite{rakovszky2020_sliom}.\\

\begin{figure}[t]
\begin{center}
\includegraphics[trim={0cm 0cm 0cm 0cm},clip,width=.99\linewidth]{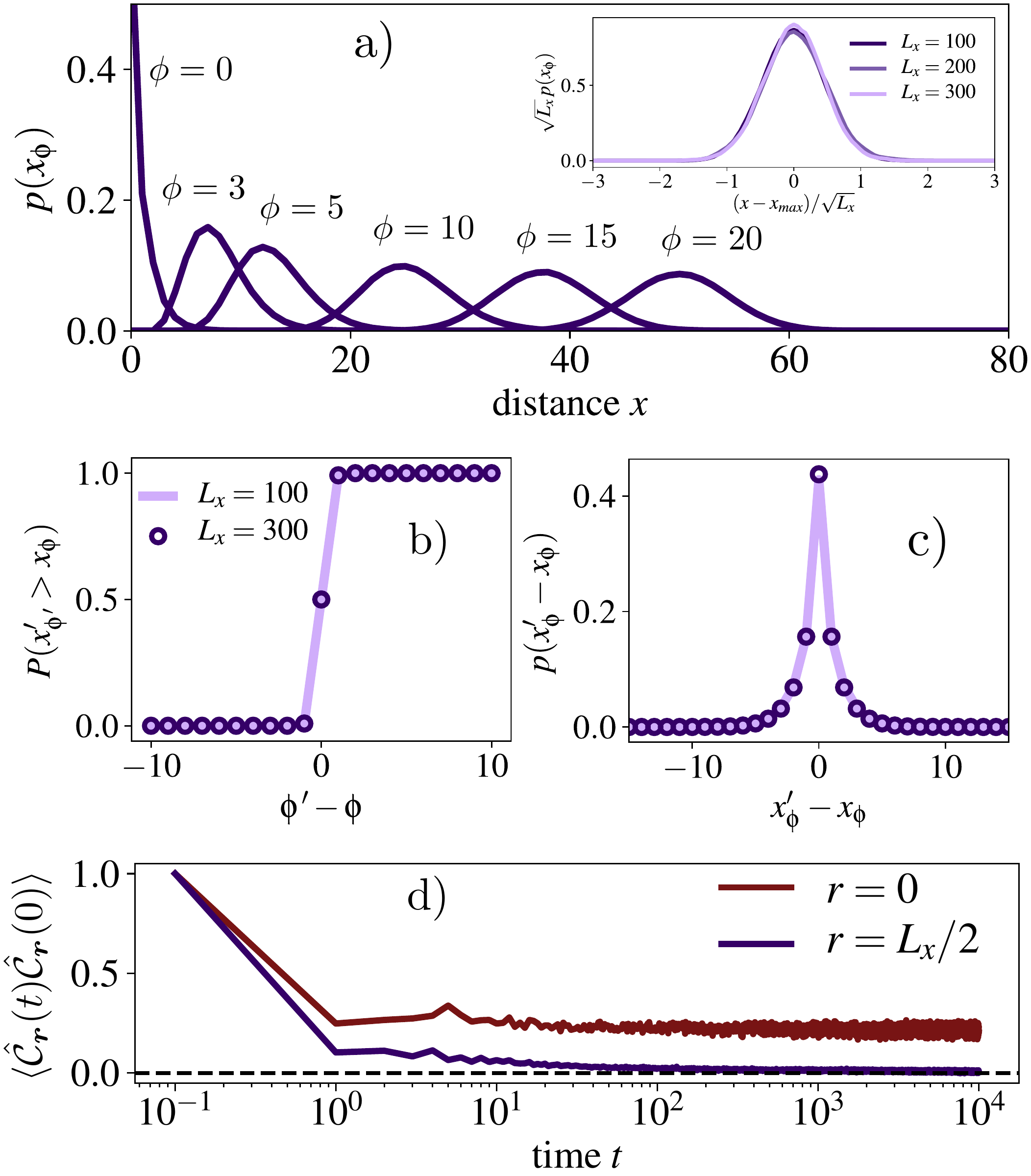}
\caption{\textbf{Statistically localized integrals of motion (SLIOMs).} \textbf{a)} Probability distribution of the $x$-value $x_\phi=r_x$ of the position of the chiral interlayer charges $\hat{q}_{\bs{r}}(\phi)$ for fixed $\phi$. The system size used for Monte Carlo simulations is $L_x=100$, $L_y=6$, as well as $W_y/L_x=0.4$. For a given $\phi$ in the bulk, the $\hat{q}_{\bs{r}}(\phi)$ are generally localized to a subextensive region of size $\sim \sqrt{L}$, see Inset. In addition, there exist modes localized exponentially on the boundary, shown here for $\phi=0$. These results demonstrate that the $\hat{Q}_\phi$ are statistically localized integrals of motion (SLIOMs) according to Ref.~\cite{rakovszky2020_sliom}. \textbf{b)} Probability of finding a chiral interlayer charge $\hat{q}_{(x^\prime_{\phi^\prime},0)}(\phi^\prime)$ to the right of another such charge $\hat{q}_{(x_\phi,0)}(\phi)$. The system-size-independent sharp step demonstrates the spatial ordering pattern of the $\hat{q}_{\bs{r}}(\phi)$ along the cylinder. \textbf{c)} Probability distribution of the distance between two interlayer charges $\hat{q}_{\bs{r}}(\phi)$ of the same $\phi$. The sharply peaked distribution shows that the SLIOMs $\hat{Q}_\phi$ can be assigned 1D positions along the cylinder. \textbf{d)} The corner charge correlation function remains finite at the boundary due to the presence of edge modes, while decaying in the bulk.}
 \label{fig:S5.2}
\end{center}
\end{figure}

To confirm this line of reasoning, we show in \figc{fig:S5.2}{a} the numerically determined probability distributions $p_\phi(x_\phi)$ of the $x_\phi=r_x$ - positions of the operators $\hat{q}_{\bs{r}}(\phi)$, for different $\phi$. More precisely, $p_\phi(x_\phi)$ is defined as
\begin{equation} \label{eq:S5.1}
p_\phi(x_\phi) := \frac{\braket{\hat{q}_{(x_\phi,0)}(\phi)}_{W_y/L_x}}{\sum_{x^\prime_\phi}\braket{\hat{q}_{(x^\prime_\phi,0)}(\phi)}_{W_y/L_x}}.
\end{equation}
We indeed find the expected localization to a $\sqrt{L_x}$ subregion within the system, for all $\phi$ scaling with system size. For $\phi$ not scaling with system size, the corresponding conserved quantities $\hat{Q}_\phi$ are instead localized close to the boundary, shown in \figc{fig:S5.2}{a} for $\phi=0$.

From \figc{fig:S5.2}{a} we also clearly see that the average positions $\sum_{x_\phi}x_\phi \, p_\phi(x_\phi) =: \overline{x_\phi} < \overline{x_{\phi+1}}$ of the chiral subcharges are spatially ordered. This spatial ordering becomes even more apparent when realizing that the probability distributions $p_\phi(x_\phi)$ for different $\phi$ are not independent: We can compute the probability distribution $\tilde{p}_{\phi^\prime,\phi}(x^\prime_{\phi^\prime}-x_{\phi})$ of the distance between two chiral interlayer charges for independent $\phi,\phi^\prime$. Formally, we define $\tilde{p}_{\phi^\prime,\phi}(x^\prime_{\phi^\prime}-x_{\phi})$ via
\begin{equation} \label{eq:S5.2}
\tilde{p}^{}_{\phi^\prime,\phi}(x^\prime_{\phi^\prime}-x_{\phi}) := \frac{\sum_{x} \braket{\hat{q}_{(x,0)}(\phi) \, \hat{q}_{(x+x^\prime_{\phi^\prime}-x_{\phi},0)}(\phi^\prime)}_{W_y/L_x}}{\sum_{x,x^\prime} \braket{\hat{q}_{(x,0)}(\phi)\, \hat{q}_{(x^\prime,0)}(\phi^\prime)}_{W_y/L_x}}.
\end{equation}
We then consider the associated probability $P_{\phi^\prime,\phi}(x^\prime_{\phi^\prime} > x_{\phi})=\sum_{x \geq 0}\tilde{p}_{\phi^\prime,\phi}(x)$ of finding the chiral interlayer charge associated to $\phi^\prime$ to the right of the one associated to $\phi$. In \figc{fig:S5.2}{b}, we see that there is a system-size-independent sharp step in $P_{\phi^\prime,\phi}(x^\prime_{\phi^\prime} > x_{\phi}) \approx \theta(\phi^\prime - \phi)$ as a function of $\phi^\prime - \phi$. Therefore, the SLIOMs $\hat{Q}_\phi$ are sharply ordered along the cylinder and thus form a conserved spatial charge pattern. Intuitively, this is understood from the fact that the $\hat{q}_{\bs{r}}(\phi)$ that contribute to $\hat{Q}_\phi$ are predominantly located between the $\phi^{th}$ and $(\phi+1)^{th}$ loop, counting from the left end of the system. In addition, the sharply peaked probability distribution $\tilde{p}_{\phi,\phi}(x^\prime_\phi-x_\phi)$ of the distance between two interlayer charges contributing to the same $\hat{Q}_\phi$ shows that in a given state, the $\hat{Q}_\phi$ can be assigned a sharp position along the cylinder, see \figc{fig:S5.2}{c}.\\

Having confirmed the presence of the statistically localized integrals of motion (SLIOMs) $\hat{Q}_\phi$, a number of results obtained in Ref.~\cite{rakovszky2020_sliom} directly carry over to our situation. First, we notice that the inversion operator $\hat{I}_z$ that exchanges the dimer configurations of upper and lower layer induces the inversion of the chirality of all loops in the projected transition graph picture. Therefore, $\{\hat{I}_z,\hat{\phi}_{\bs{r}}\}=0$ and hence also $\{\hat{I}_z,\hat{Q}_\phi\}=0$ for all $\phi$. Since also $[\hat{I}_z,\hat{H}]=0$, this implies that the spectrum of all Hilbert space sectors (except for the sector with $\hat{Q}_\phi=0$ for all $\phi$) is doubly degenerate. 

We further notice that in particular, also $\{\hat{I}_z,\hat{Q}_{\phi=0}\}=\{\hat{I}_z,\hat{Q}_{\phi=W_y-1}\}=0$, i.e. both leftmost and rightmost chiral subcharges are inverted by $\hat{I}_z$. These conserved quantities (as well as all other $\hat{Q}_\phi$ with $\phi$ not scaling with system size) are localized close to the boundary as seen in \figc{fig:S5.2}{a}, and the formal similarity to so-called strong edge modes~\cite{fendley2012_edge,fendley2016_strong,alicea2016_topo,kemp2017_coherence,else2017_prethermal,vasiloiu2019_strong} were pointed out in~\cite{rakovszky2020_sliom}. As a results of these edge modes, corner charge correlation functions at the boundary will not decay but retain a finite memory, e.g. $\lim_{t\rightarrow \infty}\braket{\hat{\mathcal{C}}_{(0,y)}(t)\, \hat{\mathcal{C}}_{(0,y)}(0)}_{W_y/L_x} \neq 0$, see \figc{fig:S5.2}{d}.

In contrast to the correlations on the boundary, the bulk corner charge correlations do decay as shown in \figc{fig:S5.2}{d}. Therefore, the strong Hilbert space fragmentation due to SLIOMs is not in general enough to prevent the system from thermalizing~\cite{rakovszky2020_sliom}. Exactly how the decay of bulk correlations ensues will be treated in the following paragraph.

\subsection{Subdiffusive relaxation}
As demonstrated in the previous section and in \figc{fig:S5.2}{d}, the bulk correlations in our cylindrical geometry decay even in the presence of a strong fragmentation of the Hilbert space due to SLIOMs. This induces the question of \textit{how} these correlations decay qualitatively, and in particular how the presence of SLIOMs influences this decay process.

To answer this question, we present a (non-rigorous) analytical argument that yields a prediction for the form of the quasi-1D correlation functions
\begin{equation} \label{eq:S5.3}
G^{(1d)}_{w_y}(x,t) := \braket{\hat{\mathcal{C}}_{(x,0)}(t)\, \hat{\mathcal{C}}_{(0,0)}(0)}_{W_y/L_x},
\end{equation}
evaluated for finite flux densities $w_y=W_y/L_x>0$. For simplicity of notation, we assume $x=0$ to be located in the bulk here. As demonstrated in the previous section, the SLIOMs $\hat{Q}_\phi$ obey a notion of locality, i.e. they form a conserved pattern of charges along the cylinder, as was similarly the case for the SLIOMs discussed in Ref.~\cite{rakovszky2020_sliom}. This pattern conservation can alternatively be interpreted as a hard core constraint, in that two different $\hat{Q}_\phi$ can never exchange relative positions along the cylinder. If $\ket{\psi}$ denotes some initial product state in the dimer occupation basis, we can then label this state $\ket{\psi} = \ket{\{Q_\phi,x_\phi\},\alpha}$ by the values of its conserved quantities $Q_\phi$ and their 1D-positions $x_\phi$ ($x_{\phi}<x_{\phi+1}$), as well as a remaining set of parameters $\alpha$ containing microscopic details. Of course, the $\hat{Q}_\phi$ are not actually site-local objects, but rather composed by all the $\hat{q}_{\bs{r}}(\phi)$. Nonetheless, from a `course-grained' point of view, we can ascribe a single $x$-position $x_\phi$ to each $Q_\phi$, see \figc{fig:S5.2}{c}. In the following, we assume the microscopic details encoded by the parameters $\alpha$ not to be essential for the transport of conserved quantities at late times, thus omiting them from the notation, i.e. $\ket{\psi}\simeq \ket{\{Q_\phi,x_\phi\}}$. The corner-charge operator $\hat{\mathcal{C}}_{(x,0)}$ will then be sensitive to the SLIOM $\hat{Q}_\phi$ that is located at $x_\phi=x$, i.e. we assume
\begin{equation} \label{eq:S5.4}
\hat{\mathcal{C}}_{(x,0)}\ket{\psi} \rightarrow \hat{\mathcal{C}}_{(x,0)}\ket{\{Q_\phi,x_\phi\}} \sim \sum_{\phi} \delta_{x_\phi,x} \, Q_\phi \, \ket{\{Q_\phi,x_\phi\}}.
\end{equation}
Inserting this assumption into the expression \eq{eq:S5.3} yields
\begin{equation} \label{eq:S5.5}
\begin{split}
&G^{(1d)}_{w_y}(x,t) = \sum_{\psi} \braket{\psi|\hat{\mathcal{C}}_{(x,0)}(t)\, \hat{\mathcal{C}}_{(0,0)}(0)|\psi}_{W_y/L_x}  \\
&\sim \sum_{\{Q_\phi,x_\phi\}} \braket{\{Q_\phi,x_\phi\}|\hat{\mathcal{C}}_{(x,0)}(t)\, \hat{\mathcal{C}}_{(0,0)}|\{Q_\phi,x_\phi\}} \\
&\sim \sum_{\{Q_\phi,x_\phi\}} \sum_\phi \delta_{x_\phi,0} \, Q_\phi \braket{\{Q_\phi,x_\phi(t)\}|\hat{\mathcal{C}}_{(x,0)}|\{Q_\phi,x_\phi(t)\}} \\
&\sim \sum_{\{Q_\phi,x_\phi (0)\}} \sum_{\phi,\phi^\prime} \delta^{}_{x_\phi(0),0} \, \delta_{x_{\phi^\prime}(t),x} \, Q_\phi \, Q_{\phi^\prime},
\end{split}
\end{equation}
where we have substituted $\sum_\psi \rightarrow \sum_{\{Q_\phi,x_\phi\}}$ and shifted the time dependence to the states in the last two lines. We now assume further that the position $x_\phi(t)$ of $\hat{Q}_\phi$ at time $t$ does not depend on the value of $Q_\phi$, thus again neglecting certain microscopic details. We can then directly carry out the average $\sum_{\{Q_\phi\}} Q_\phi Q_{\phi^\prime} \sim \delta_{\phi,\phi^\prime}$ to obtain
\begin{equation} \label{eq:S5.6}
G^{(1d)}_{w_y}(x,t) \sim \sum_{\{x_\phi (0)\}} \sum_{\phi} \delta_{x_\phi(0),0}\, \delta_{x_{\phi}(t),x}.
\end{equation}
\eq{eq:S5.6} has an intuitive interpretation: $G^{(1d)}_{w_y}(x,t)$ describes the tracer motion of individual SLIOMs $\hat{Q}_\phi$, which move from $x_\phi(0)=0$ at time $t=0$ to $x_\phi(t)=x$ at time $t$.  Notice that the $\hat{Q}_\phi$ become effectively \textit{distinguishable} particles due to the (initial state) average $\sum_{\{Q_\phi\}}$.

\begin{figure}[t]
\begin{center}
\includegraphics[trim={0cm 0cm 0cm 0cm},clip,width=.9\linewidth]{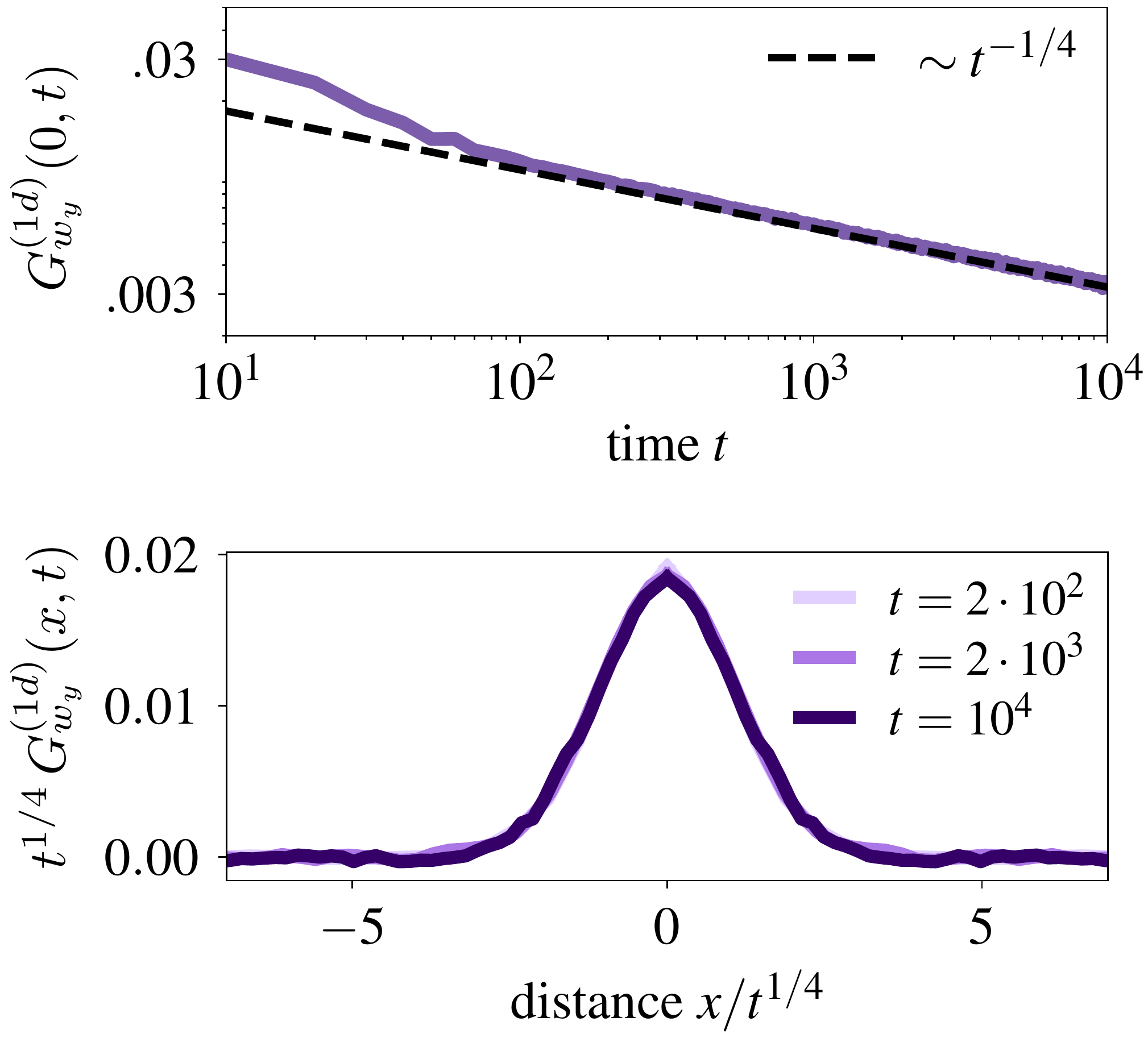}
\caption{\textbf{Relaxation of corner charges in 1D.} Local correlations of the corner charge relax subdiffusively in an effective 1D geometry (cylinder with finite circumference). The exponent of the decay is $1/4$ and the correlations assume a Gaussian form. This is in full agreement with the probabilty distributions of hard-core tracer particles, expected to describe the late time decay of SLIOM correlation functions.}
 \label{fig:S5.3}
\end{center}
\end{figure}

Recalling that the $\hat{Q}_\phi$ obey an effective hard core constraint, we recognize that due to \eq{eq:S5.6}, the motion of SLIOMs should effectively be described by the tracer diffusion of hard core particles in one dimension. This problem has been studied within more direct setups in the mathematical literature and admits an exact solution for the asymptotic probability distribution of hard core tracer particles at long times~\cite{harris1965_diffusion,levitt1973_dynamics,vanBeijeren1983_diffusion} (see in particular Ref.~\cite{vanBeijeren1983_diffusion} and references therein for an overview of the history of this problem). This probability distribution directly carries over to the correlations $G^{(1d)}_{w_y}(x,t)$ via \eq{eq:S5.6}, and we thus predict
\begin{equation} \label{eq:S5.7}
G^{(1d)}_{w_y}(x,t) = (Dt)^{-1/4}\, \exp(-x^2/\sqrt{Dt}),
\end{equation}
for the long time hydrodynamic decay of correlations in systems hosting SLIOMs in general, and our bilayer dimer setup specifically. For the latter, we can immediately verify the validity of \eq{eq:S5.7} numerically, as shown in \fig{fig:S5.3}. Notice that the correlations assume a Gaussian shape, but decay \textit{subdiffusively} slow, with $G^{(1d)}_{w_y}(0,t) \sim t^{-1/4}$ for the return probability (cf. $G(0,t)\sim t^{-1/2}$ for normal diffusion in 1D).

To conclude this section, we note that we expect the result \eq{eq:S5.7} to be an \textit{a priori} consequence of the presence of SLIOMs in arbitrary systems under a sufficiently ergodic time evolution. While we have explicitly used in \eq{eq:S5.5} that at each point in time, the system is in a product state in the automaton evolution, a similar reasoning in terms of hard core tracer diffusion should apply equally well for any generic plaquette dynamics. It would be interesting to verify this prediction explicitly in the future, e.g. for systems such as the $t-J_z$-model discussed in Ref.~\cite{rakovszky2020_sliom}.

\section{Connections to Topological Solitons} \label{sec:solitons}
As announced above in Sec.~\ref{sec:2}, the global conserved quantity $\hat{\mathcal{Q}}$, whose exotic associated transport properties we have investigated in the main body of this work, can be interpreted as a topological soliton conservation law. More specifically, we will show in this section that the total chiral charge $\hat{\mathcal{Q}}$ corresponds to the bilayer version of a conserved \textit{Hopf-invariant} that exists more generally in the cubic lattice dimer model as derived in Refs.~\cite{freedman_hopfions,bednik_hopfions}. The correlation functions we considered previously can thus be interpreted as characterizing the dynamics of \textit{Hopfions} (i.e. three-dimensional topological solitons) within the bilayer geometry.

\begin{center}
\textit{Hopfions: A brief introduction}
\end{center}

Before specifically analyzing the abovementioned reformulation of $\hat{\mathcal{Q}}$ as a conserved Hopf-number, let us first take a small detour to introduce the concept of Hopfions more generally: Hopfions are three-dimensional topological solitons, originally introduced in Ref.\cite{Hopf_1964}. They can be defined in terms of the homotopy classes of maps between 3- and  2-spheres, $\bs{n}: \mathcal{S}^3 \rightarrow \mathcal{S}^2$. As $\mathcal{S}^3$ is isomorphic to $\mathbb{R}^3 \cup \{\infty\}$ by stereographic projection, we can think of $\bs{n}(\bs{r})$ as a unit vector field in $\mathbb{R}^3$ with a uniform limit $\bs{n}(|\bs{r}|\rightarrow \infty) = \bs{n}_0$. The fibres of this vector field, defined as the preimages $\bs{n}^{-1}(\bs{q}) \subset \mathbb{R}^3$ of given points $\bs{q}\in \mathcal{S}^2$ on the 2-sphere, form closed loops in $\mathbb{R}^3$. The linkage number $N_\mathcal{H}(\bs{n})$ of two such fibres under the map $\bs{n}$ yields the directed number of times two such loops are winding around each other, thereby providing an integer homotopy classification of $\bs{n}$. Within this interpretation of linking numbers of preimages, the necessity of a three-dimensional setting in order to provide a non-trivial Hopf-invariant is evident.

For practical computational purposes, the Hopf invariant $N_\mathcal{H}(\bs{n})$ can be expressed as an integral,
\begin{equation} \label{eq:1.1}
N_\mathcal{H}(n) = \int d^3r \,\bs{B}(\bs{r})\cdot \bs{A}(\bs{r}),
\end{equation}
where the 'magnetic field' is given by $B_i=\epsilon_{ijk}\,\bs{n}\cdot (\nabla_j\bs{n}\times \nabla_k \bs{n})$ and the implicit vector potential $\bs{B}=\nabla \times \bs{A}$. The expression of Eq.\eqref{eq:1.1} is typically applied to classical field theories where $\bs{n}(\bs{r})$ can be interpreted as a magnetization vector field in a solid state system. As opposed to their two-dimensional Skyrmion counterparts~\cite{roessler2006_skyrmion,muehlbauer2009_skyrmion}, the stabilization of magnetic configurations with non-trivial Hopf-numbers have so far eluded experimental detection in solid state systems, and are subject to active research also in the context of topological phases of matter~\cite{moore2008_hopf,deng2013_hopf,liu2017_hopfSPT}.

\begin{center}
\textit{Hopfions in the dimer model}
\end{center}

For the present lattice dimer model, the connection to the Hopf invariant of Eq.\eqref{eq:1.1} can be made in two ways: Either by a suitable continuum limit which allows for a direct use of Eq.\eqref{eq:1.1} \cite{freedman_hopfions}. Or, by providing a discrete lattice version of the invariant Eq.\eqref{eq:1.1} \cite{bednik_hopfions}, which is the approach we will adopt in the following. We emphasize that in order to define a Hopf number, we have to assume OBCs (therefore, in the end, the quantity $\hat{\mathcal{Q}}$ reduces to the Hopf number upon choosing open boundaries).

\begin{figure}[t]
\begin{center}
\includegraphics[trim={0cm 0cm 0cm 0cm},clip,width=.99\linewidth]{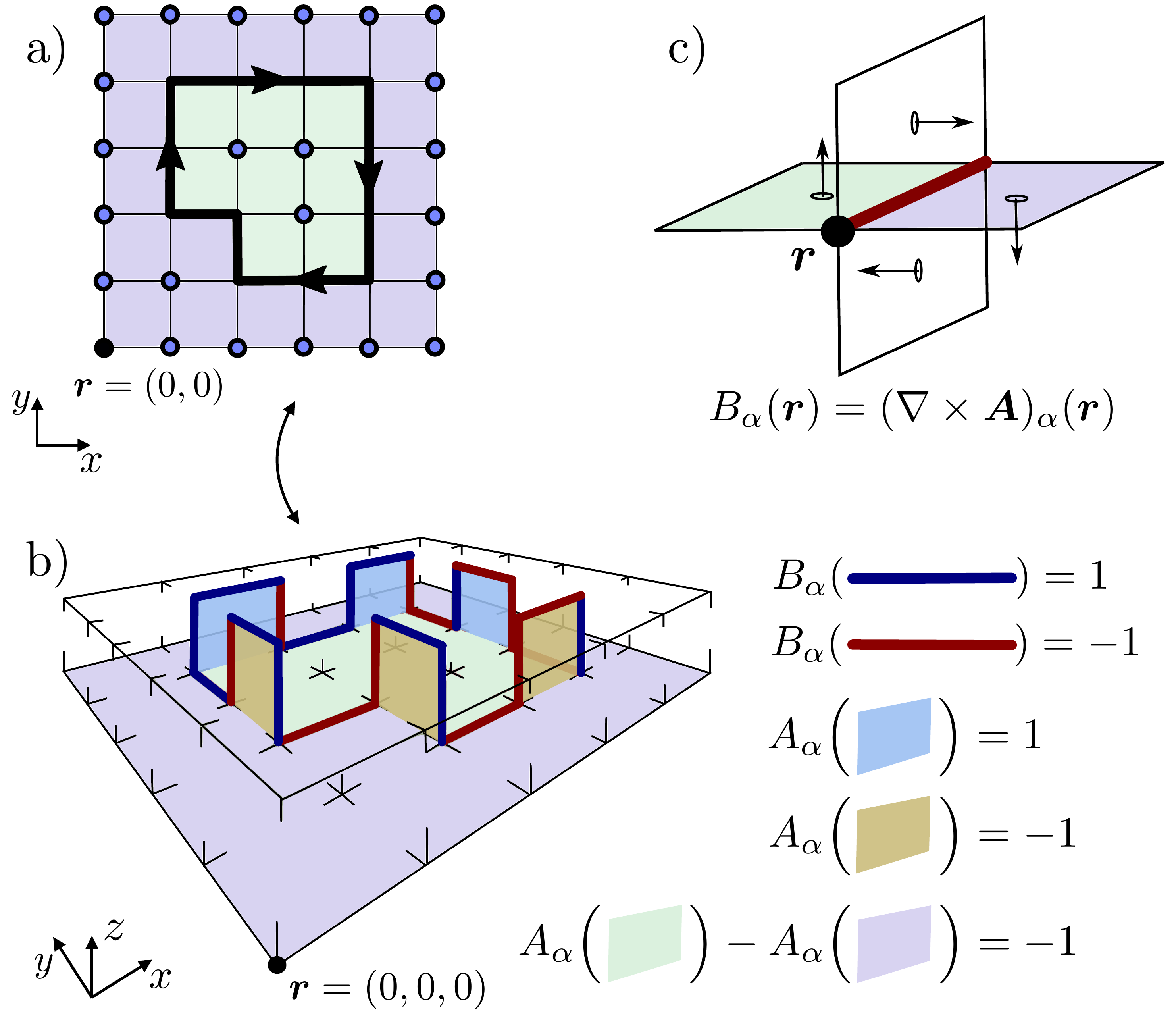}
\caption{\textbf{Lattice magnetic field description.} A given dimer configuration containing a certain loop in its transition graph in a) can be mapped to a lattice magnetic field $B_\alpha(\bs{r})$ on the bonds of the bilayer lattice according to \eq{eq:1.2} in b). The magnetic field can be derived from an associated vector potential $A_\alpha(\bs{r})$ living on the plaquettes of the lattice, see c). The values of the vector potential for the example in b) where chosen according to \eq{eq:S6.2} and \eq{eq:S6.3}. It can directly be verified that this choice leads to the correct dimer occupation numbers, as well as a Hopf number $N_\mathcal{H}$ from \eq{eq:1.6} that agrees with the total chiral charge $\hat{\mathcal{Q}}$ from \eq{eq:S3.5}, evaluated for the loop in a).}
\label{fig:S6.1}
\end{center}
\end{figure}

Following Refs.~\cite{Huse2003_coulomb,bednik_hopfions}, we first have to choose a lattice magnetic field description of our dimer model. For this purpose, we define a field on the bonds of the lattice,
\begin{equation} \label{eq:1.2}
B_\alpha(\bs{r}) = (-1)^{r_x+r_y+r_z}\left[\hat{n}^{(d)}_{\bs{r},\alpha}-\delta^{}_{\alpha,z}\,\delta_{(-1)^{r_z},1}\right],
\end{equation}
with $\alpha \in \{x,y,z\}$, that can be verified to satisfy a zero divergence condition
\begin{equation} \label{eq:1.3}
\nabla \cdot \bs{B}(\bs{r}) = \sum_\alpha \left[B_\alpha(\bs{r})-B_\alpha(\bs{r}-\bs{e}_\alpha)\right] = 0,
\end{equation}
see \fig{fig:S6.1} for an example. Using \eq{eq:1.2}, every dimer configuration maps uniquely to a magnetic field configuration.
We can then think of our $L_x\times L_y\times 2$ bilayer-system with OBCs as being embedded within an infinite cubic lattice. Outside the bilayer system we fix the dimers to a trivial configuration $\hat{n}^{(d)}_{\bs{r},\alpha} = \delta_{\alpha,z}\,\delta_{(-1)^{r_z},1}$ for $\bs{r} \notin [0,L_x]\times[0,L_y]\times[0,1]$, which implies a vanishing magnetic field $\bs{B}(\bs{r})=0$ on all bonds not part of the finite bilayer system. Notice that this property is consistent with the condition $\bs{n}(|\bs{r}|\rightarrow \infty)=const.$ required in the usual continuum definition of the Hopf number mentioned above.

With a magnetic field living on the \textit{bonds} $(\bs{r},\alpha)$ of the lattice at hand, the associated discrete vector potential $\bs{A}(\bs{r})$ is defined on its \textit{plaquettes}. If  $A_\gamma(\bs{r})$ denotes the vector potential on the plaquette whose center lies at $\bs{r}+\bs{e}_\alpha/2+\bs{e}_\beta/2$ ($(\alpha,\beta,\gamma) \in \mathrm{Perm}(x,y,z)$) with normal vector $\bs{e}_\gamma$, the relation between magnetic field and vector potential can be expressed as
\begin{equation} \label{eq:1.5}
B_\alpha(\bs{r}) = \left(\nabla \times \bs{A}(\bs{r})\right)_\alpha = \epsilon_{\alpha\beta\gamma} \left[  A_\gamma(\bs{r})-A_\gamma(\bs{r}-\bs{e}_\beta) \right].
\end{equation}
Hence, once the values of the vector potential are known, the corresponding magnetic field values can simply be determined via a 'right-hand-rule', see \figc{fig:S6.1}{c} for an illustration.\\

Equipped with these lattice definitions, the corresponding discrete equivalent to the Hopf number Eq.\eqref{eq:1.1} for a given dimer configuration was given in Ref.\cite{bednik_hopfions} as
\begin{equation} \label{eq:1.6}
\begin{split}
&N_\mathcal{H} = \frac{1}{8}\sum_{\bs{r}} \bs{A}(\bs{r})\cdot \bar{\bs{B}}(\bs{r}) = \frac{1}{8}\sum_{\bs{r},\alpha} A_\alpha(\bs{r})\, \bar{B}_\alpha(\bs{r})=\\
&= \frac{1}{8}\sum_{\bs{r},\alpha} A_\alpha(\bs{r})\Bigl[ B_\alpha(\bs{r})+B_\alpha(\bs{r}+\bs{e}_\beta)+B_\alpha(\bs{r}+\bs{e}_\gamma)+\\
&+ B_\alpha(\bs{r}+\bs{e}_\beta+\bs{e}_\gamma) + B_\alpha(\bs{r}-\bs{e}_\alpha) +B_\alpha(\bs{r}-\bs{e}_\alpha+\bs{e}_\beta) +\\
&+ B_\alpha(\bs{r}-\bs{e}_\alpha+\bs{e}_\gamma) + B_\alpha(\bs{r}-\bs{e}_\alpha + \bs{e}_\beta  + \bs{e}_\gamma) \Bigr],
\end{split}
\end{equation}
where the term $\bar{B}_\alpha(\bs{r})$ in brackets can be considered as the average magnetic field adjacent to the plaquette $A_\alpha (\bs{r})$, providing an analogy to the form of Eq.\eqref{eq:1.1}. The invariance of Eq.\eqref{eq:1.6} either under gauge transformations of the vector potential,
\begin{equation} \label{eq:1.7}
A_\alpha(\bs{r}) \rightarrow A_\alpha(\bs{r}) + f(\bs{r})-f(\bs{r}-\bs{e}_\alpha)
\end{equation}
with some scalar function $f$, leaving the magnetic field invariant, as well as under elementary plaquette flips
\begin{equation} \label{eq:1.8}
A_\alpha(\bs{r}) \rightarrow A_\alpha(\bs{r}) \pm 1
\end{equation} 
with a correspondingly transforming $B$-field according to Eq.\eqref{eq:1.5}, was demonstrated in Ref.\cite{bednik_hopfions}.

\begin{center}
\textit{Hopf-charge of conserved quantities}
\end{center}

To show that in the bilayer geometry, $N_\mathcal{H}$ is indeed the total chiral charge $\mathcal{Q}$ from \eq{eq:S3.5}, we first recognize that according to \eq{eq:1.2}, the magnetic field is non-zero only on bonds that are part of loops within the transition graph picture (as well as on the interlayer $z$-bonds along such loops), see \fig{fig:S6.1}{b} for an illustration. If we characterize a certain dimer configuration via the collection $\{\mathcal{L}\}$ of loops contained within its transition graph, we can show that $N_\mathcal{H}(\{\mathcal{L}\}) = \sum_{\mathcal{L} \in \{\mathcal{L}\}} N_\mathcal{H}(\mathcal{L})$ can be expressed as a sum over the Hopf numbers of individual loops: If a vector potential $\bs{A}_{1/2}(\bs{r})$ and its induced field $\bs{B}_{1/2}(\bs{r})$ lead to the (non-overlapping) loops $\mathcal{L}_{1/2}$ in the transition graph respectively, then $\bs{A}(\bs{r})=\bs{A}_1(\bs{r})+\bs{A}_2(\bs{r})$ contains both $\{\mathcal{L}\}=\{\mathcal{L}_1,\mathcal{L}_2\}$ in its induced transition graph. According to \eq{eq:1.6}, the Hopf number of the dimer configuration described by $\bs{A}(\bs{r})$ is then given by 
\begin{equation} \label{eq:S6.1}
\begin{split}
&N_\mathcal{H}(\{\mathcal{L}_1,\mathcal{L}_2\}) = N_\mathcal{H}(\mathcal{L}_1) + N_\mathcal{H}(\mathcal{L}_2) +\\
&\quad + \frac{1}{8}\sum_{\bs{r}} \bs{A}_1(\bs{r})\cdot \bar{\bs{B}}_2(\bs{r}) + \frac{1}{8}\sum_{\bs{r}} \bs{A}_2(\bs{r})\cdot \bar{\bs{B}}_1(\bs{r}).
\end{split}
\end{equation}
Since the two loops are non-overlapping by virtue of the hard core constraint, $\bs{A}_{1/2}(\bs{r})$ does not generate a finite field strength on all $\bs{r}$ where $\bs{B}_{2/1}(\bs{r})\neq 0$ is finite. Hence, by application of a suitable gauge transformation \eq{eq:1.7}, $\bs{A}_{1/2}(\bs{r})=0$ can be chosen to vanish on all such $\bs{r}$ and the cross-terms in the second line of \eq{eq:S6.1} indeed vanish. 
Therefore, we only have to show that $N_\mathcal{H}(\mathcal{L}) = \mathcal{Q}(\mathcal{L})$ for dimer configurations containing a single loop $\mathcal{L}$ (all other dimers are then fixed along $z$-bonds) in their transition graph. This is done most easily by providing a specific instance of a vector potential $\bs{A}(\bs{r})$ that produces the loop $\mathcal{L}$ in the transition graph, and subsequently inserting this $\bs{A}(\bs{r})$ into \eq{eq:1.6}. 

To achieve this, let us first denote by $p=p(\bs{r})=\{\bs{r},\bs{r}+\bs{e}_x,\bs{r}+\bs{e}_x+\bs{e}_y,\bs{r}+\bs{e}_y\}$ the set that contains the four sites of an elementary plaquette on the 2D square lattice. Recall that a given loop $\mathcal{L}=\{\bs{r}_0,...,\bs{r}_{|\mathcal{L}|-1}\}$ is given by an ordered set of sites and the direction of the ingoing loop segment at $\bs{r}_n$ is given by $\bs{\ell}_i(\bs{r}_n) = \bs{r}_{n}-\bs{r}_{n-1}$. Recall further, that $v_\mathcal{L}$ denotes the interior of the 2D loop $\mathcal{L}$, see \fig{fig:3} and the discussion in Appendix~\ref{sec:app2}.
Then, the following vector potential will lead to a configuration that contains the loop $\mathcal{L}$ in its transition graph:
\begin{itemize}
\item For all $\bs{r},\bs{r}^\prime \in \mathbb{Z}^2$ s.t. $p(\bs{r})\cap v_\mathcal{L} = \emptyset$ and $p(\bs{r}^\prime)\cap v_\mathcal{L} \neq \emptyset$, choose
\begin{equation} \label{eq:S6.2}
\begin{split}
A_z(r_x,r_y,0) - A_z(r_x^\prime,r_y^\prime,0) = 1.
\end{split}
\end{equation}

\item For all $\bs{r}_n$ on the $A$-sublattice and $\alpha \in \{x,y\}$ chosen such that $\bs{e}_\alpha \cdot \bs{\ell}_i(\bs{r}_n)=0$, choose
\begin{equation} \label{eq:S6.3}
\begin{split}
A_\alpha\Bigl(\bs{r}_n-\bs{\ell}_i(\bs{r}_n) \frac{1+\bs{\ell}_i(\bs{r}_n)\cdot (\bs{e}_x+\bs{e}_y)}{2} \Bigr) = \\
= (\bs{e}_x+\bs{e}_y) \cdot [\bs{\ell}_i(\bs{r}_n) \times (\bs{e}_x+\bs{e}_y)].
\end{split}
\end{equation}

\item Choose $A_\alpha(\bs{r})=0$ for all remaining plaquettes.

\end{itemize}

Using \eq{eq:1.2} and \eq{eq:1.5}, it is straightforward to check that this choice of $\bs{A}(\bs{r})$ yields the correct dimer configuration that produces the loop $\mathcal{L}$. Furthermore, inserting \eq{eq:S6.2} and \eq{eq:S6.3} into the expression \eq{eq:1.6} for the Hopf Number $N_\mathcal{H}$, a lengthy but straightforward calculation shows that indeed $N_\mathcal{H}=\hat{\mathcal{Q}}$ is the same as the total chiral charge conservation law of \eq{eq:S3.5}. It is very instructive to convince oneself of the validity of $N_\mathcal{H}=\hat{\mathcal{Q}}$ through \fig{fig:S6.1}. In this \fig{fig:S6.1}{b}, we have entered the values of $\bs{A}(\bs{r})$ according to \eq{eq:S6.2} and \eq{eq:S6.3} for a specific example. The associated magnetic field values, dimer occupation numbers, and the Hopf number can then directly be read off.

As a results of these considerations, we conclude that the fractonic corner charges $\hat{\mathcal{C}}_{\bs{r}}$ carry a non-vanishing Hopf-charge. Similarly, the conserved chiral subcharges $\hat{Q}_\phi$ can be viewed as independently conserved Hopfion-subcharges, which provides an intriguing interpretation for the dynamics studied in Sec.~\ref{sec:fractons} and Sec.~\ref{sec:q1D}. Since the Hopf charge exists also on the fully three-dimensional cubic lattice, it would be interesting to study which of our observed features, and under what circumstances, might carry over higher dimensions.

\section{Conclusion and Outlook}
In this work we have investigated the non-equilibrium properties of a bilayer dimer model using classically simulable automata circuits, adding to increasing recent interest in the dynamics of dimer models~\cite{oakes2016_coops,lan2017_dimereth,lan2018_glassy,feldmeier2019_glassy,theveniaut2020_dimermbl,
pietracaprina2020_dimermbl,flicker2020_penrose}. We have found fracton-like dynamics of objects we termed corner charges that are associated to a globally conserved chiral charge, which we have found to be equivalent to a topological soliton conservation law. The dynamics of the full quasi-2D system for finite flux densities is characterized by the formation of effective one-dimensional tubes that restrict the mobility of corner charges, a hallmark of fractonic behavior. This leads to an anomalously slow decay of local correlations, as charges can diffusive only along one instead of two independent directions. Since the 1D tubes can only be destabilized by moving non-local winding loops through the system, they are stable up to a time that appears to diverge with system size, leading to non-ergodic behavior in the thermodynamic limit. In addition, we have identified the presence of statistically localized integrals of motion (SLIOMs) in a quasi-one-dimensional limit of the model. The hydrodynamic relaxation of these SLIOMs was found to be subdiffusively slow and can be described by the tracer diffusion of classical hard core particles.
The applicability of this latter result extends beyond the specific model studied in this paper and describes the hydrodynamic behavior of SLIOMs more generally -- provided they are not so strong as to localize the system as in Ref.~\cite{Sala2020_ergodicity}. In particular, verifying this expectation for systems like the $t$-$J_z$-model, whose SLIOMs were derived in Ref.~\cite{rakovszky2020_sliom} and for which a closed-system quantum time evolution is numerically feasible, is an interesting prospect for future study.

Moreover, the results derived in this work should apply to bilayer versions of arbitrary dimer models on bipartite planar lattices, for which most of our constructions are expected to proceed in an analogous way. Which of our results and under what circumstances might also generalize to dimer models in the fully three-dimensional limit is less apparent. In particular, while the soliton conservation law utilized in this work exists in the 3D cubic dimer model as well, the equivalent of corner charges and the effect of finite flux densities is left as an open question.

Other than changing the lattice geometry, we can also vary the underlying static electric charge distribution of the lattice gauge theory that is dual to the dimer model~\cite{Celi2020_2Dgauge}. Potential future work might conduct a systematic survey on the presence of soliton conservation laws depending on the underlying charge distribution. This could open a window for a more general glimpse into the thermalization dynamics of gauge theories via the study of late time transport properties.

Finally, while proposals to study lattice gauge theories like dimer models experimentally with Rydberg quantum simulators have already been put forward in~\cite{Celi2020_2Dgauge} (although for the planar 2D case), it will be interesting to see whether bilayer dimer models can also potentially be obtained as realistic low energy theories in condensed matter systems such as (artificial) spin ice, or in the strong coupling limit of correlated fermion models~\cite{pollmann2006_corr}. Naturally, interest then also extends towards the equilibrium properties of such models~\cite{wilkins2020_double,desai2020_bilayer}.\\

\textbf{\textit{Acknowledgments.}}--
We thank Tibor Rakovszky and Pablo Sala for many insightful discussions.
We acknowledge support from the Technical University of Munich - Institute for Advanced Study, funded by the German Excellence Initiative and the European Union FP7 under grant agreement 291763, the Deutsche Forschungsgemeinschaft (DFG, German Research Foundation) under Germanys Excellence Strategy–EXC–2111–390814868, Research Unit FOR 1807 through grants No. PO 1370/2-1, TRR80 and DFG grant No. KN1254/2-1, No. KN1254/1-2, and from the European Research Council (ERC) under the European Unions Horizon 2020 research and innovation programme (grant agreements No. 771537 and No. 851161).\\

\appendix

\section{Proof of \eq{eq:S3.4}} \label{sec:app3}
We first restate \eq{eq:S3.4} of the main text in a more formal version in order to set up the proof.\\

\textit{Claim:} Let us consider an arbitrary directed loop $\mathcal{L}=\{\bs{r}_0,\bs{r}_1,...,\bs{r}_{N-1}\}$ on the square lattice that fulfills the following two conditions:
\begin{enumerate}[label=(\roman*)]
\item $\mathcal{L}$ is \textit{closed:} $|\bs{r}_{n+1}-\bs{r}_n| = 1$ for all $n\in \{0,...,N-1\}$, with $\bs{r}_N \equiv \bs{r}_0$.
\item $\mathcal{L}$ is \textit{non-intersecting:} $\bs{r}_n = \bs{r}_{m} \iff n=m$.
\end{enumerate}
Furthermore, let us denote by $v_{\mathcal{L}}\subset \mathbb{Z}^2$ the set of lattice points that form the interior of the loop $\mathcal{L}$ as shown in \figc{fig:3}{a} (see also a more formal definition of $v_\mathcal{L}$ in the discussion around \eq{eq:Abb.2} of Appendix~\ref{sec:app2}).\\

Given these definitions, the following identity holds:
\begin{equation} \label{eq:A.b1}
\Delta N_{AB}(v_{\mathcal{L}}) = \frac{1}{4} \sum_{n=0}^{N-1} (-1)^{x_n+y_n}\, \Bigl( \bs{\ell}^{}_{o}(\bs{r}_n) \wedge \bs{\ell}^{}_{i}(\bs{r}_n) \Bigr),
\end{equation}
where $\Delta N_{AB}(v_{\mathcal{L}})$ is the difference between the number of $A/B$ sublattice sites contained within the set $v_{\mathcal{L}}$, and $\bs{\ell}^{}_{o}(\bs{r}_n)=\bs{r}_{n+1}-\bs{r}_n$, $\bs{\ell}^{}_{i}(\bs{r}_n)=\bs{r}_{n}-\bs{r}_{n-1}$. Note that $\bs{\ell}^{}_{i/o} \in \{\pm \bs{e}_x, \pm \bs{e}_y\}$, and $\bs{r}_{-1} = \bs{r}_0$ by definition. The symbol `$\wedge$' denotes the wedge-product, which yields a scalar for the two-dimensional vectors considered here: $\bs{a}\wedge \bs{b} = a_xb_y-a_yb_x$.\\

\begin{figure}[t]
\begin{center}
\includegraphics[trim={0cm 0cm 0cm 0cm},clip,width=.99\linewidth]{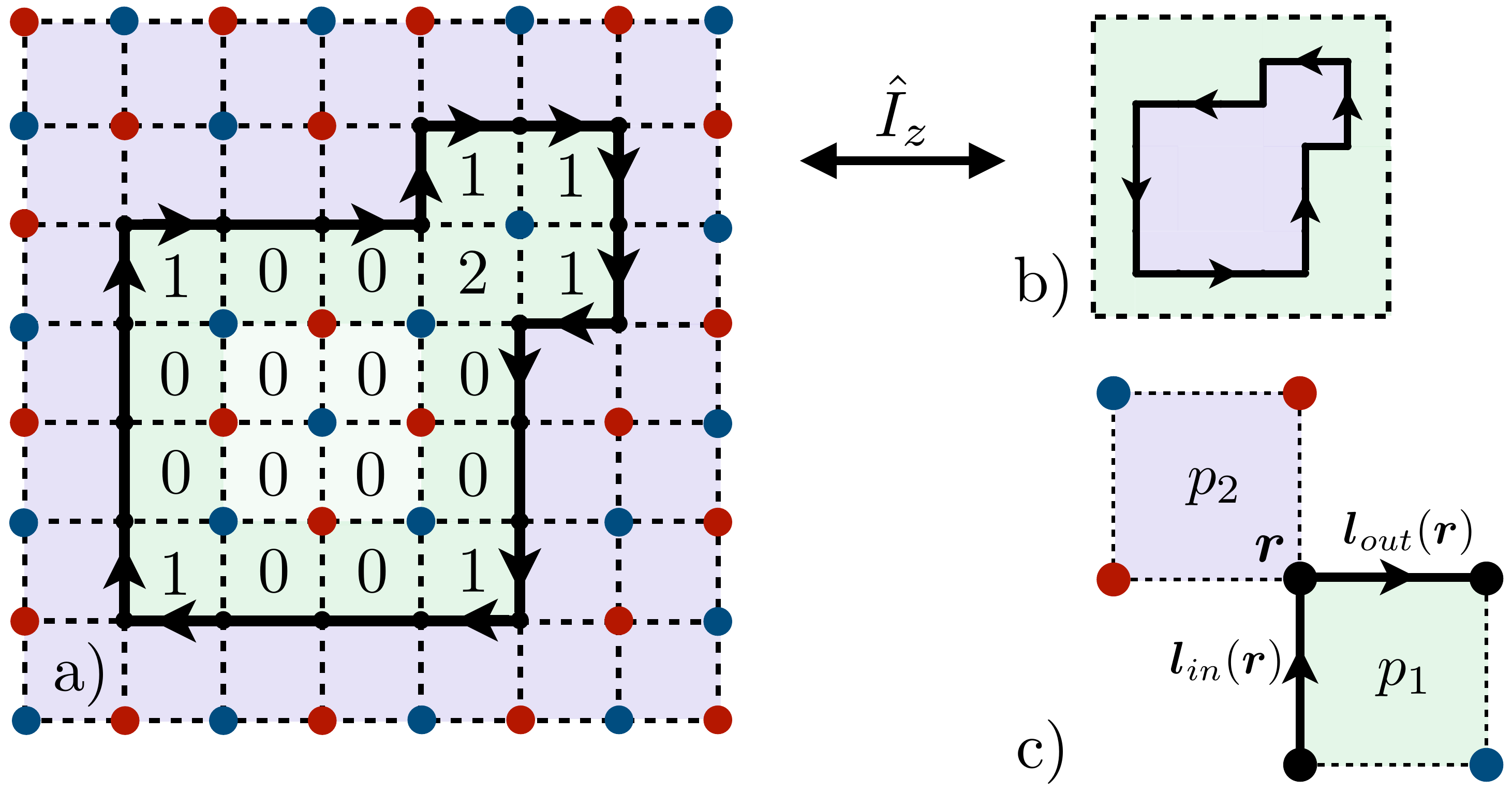}
\caption{\textbf{Proof of \eq{eq:S3.4}.} \textbf{a)} A closed, directed loop $\mathcal{L}$ (black) encloses a region $V_{\mathcal{L}}$ (green shaded). The difference in the number of $A$ (blue) and $B$ (red) sublattice sites contained within $V_{\mathcal{L}}$ can be traced back to the corners of $\mathcal{L}$. \textbf{b)} Changing the direction of $\mathcal{L}$ exchanges interior and exterior of $\mathcal{L}$. \textbf{c)} For every corner, there exist two plaquettes $p_1$ and $p_2$ that potentially obtain a non-trivial difference of $A$ and $B$ sublattice sites. Which of these plaquettes is contained within the area $V_{\mathcal{L}}$ enclosed by $\mathcal{L}$ is determined by the chirality of $\mathcal{L}$.}
\label{fig:A.b1}
\end{center}
\end{figure}

\textit{Proof:}
Let us first denote by 
\begin{equation} \label{eq:A.b2}
p(\bs{r}) = \{\bs{r},\bs{r}+\bs{e}_x,\bs{r}+\bs{e}_x+\bs{e}_y,\bs{r}+\bs{e}_y\}
\end{equation}
the four sites contained within an elementary plaquette $p$ of the square lattice. Since 
\begin{equation} \label{eq:A.b22}
\bigcup_{p\, \subseteq (v_{\mathcal{L}} \cup \mathcal{L}) }^{} p\backslash \mathcal{L} = v_{\mathcal{L}},
\end{equation} 
we can rewrite the left hand side of \eq{eq:A.b1} as
\begin{equation} \label{eq:A.b3}
\Delta N_{AB}(v_{\mathcal{L}}) = \frac{1}{4} \sum_{p\, \subseteq (v_{\mathcal{L}} \cup \mathcal{L}) }  \Delta N_{AB}\bigl(p\backslash \mathcal{L}\bigr),
\end{equation}
with the sum running over all plaquettes contained within $v_{\mathcal{L}} \cup \mathcal{L}$. The factor $1/4$ compensates for the overcounting that results from each site being adjacent to four different plaquettes, see \figc{fig:A.b1}{a}.

We see that for those plaquettes in $v_{\mathcal{L}} \cup \mathcal{L}$ that are not touched by $\mathcal{L}$, i.e. $p\backslash \mathcal{L} = p$, we have $\Delta N_{AB}(p)= 2-2 =0$ immediately, see \figc{fig:A.b1}{a}. Thus, it is sufficient to focus on plaquettes with a non-vanishing intersection $p\cap \mathcal{L} \neq \emptyset$. Crucially, we then recognize that a non-vanishing $\Delta N_{AB}\bigl(p\backslash \mathcal{L}\bigr) \neq 0$ can only be realized if there is at least one connected section of $\mathcal{L}$ passing through an \textit{odd} number of sites in $p$, see \figc{fig:A.b1}{c}.
We can then expand
\begin{widetext}
\begin{equation} \label{eq:A.b4}
\begin{split}
\Delta N_{AB}(p\backslash \mathcal{L}) &= \sum_{\bs{r} \, \in \, p\backslash \mathcal{L}} (-1)^{r_x+r_y} = - \sum_{\bs{r} \, \in \, p \, \cap \, \mathcal{L}} (-1)^{r_x+r_y} = -\sum_{n=0}^{N-1} (-1)^{x_n+y_n} \bigl| \{\bs{r}_n\}\, \cap \, p \bigr| = \\
&= \sum_{n = 0}^{N-1} (-1)^{x_n+y_n} \bigl| \{\bs{r}_n\}\, \cap \, p \bigr|\, \Bigl[ \bigl| \left\{\bs{r}_{n+1} \right\} \cap p \bigr| \bigl| \left\{\bs{r}_{n-1} \right\} \cap p \bigr| - \left(1-\bigl| \left\{\bs{r}_{n+1}\right\} \cap p \bigr|\right) \left(1-\bigl| \left\{\bs{r}_{n-1}\right\} \cap p \bigr| \right)\Bigr],
\end{split}
\end{equation}
\end{widetext}
where $|...|$ denotes the number of elements contained within a given set, and $\bigl| \left\{ \bs{r} \right\} \cap p \bigr| \, \in \, \{0,1\}$ determines whether $\bs{r}$ is contained in $p$ or not. The first term in the square brackets of \eq{eq:A.b4} corresponds to a section $\{\bs{r}_{n-1},\bs{r}_n,\bs{r}_{n+1}\} \subset \mathcal{L}$ running through three sites of the plaquette $p$, while in the second term only the site $\bs{r}_n$, and \textit{not} the sites $\bs{r}_{n-1},\bs{r}_{n+1}$, is part of $p$. These contributions correspond to $p_1$ and $p_2$ in \figc{fig:A.b1}{c}, respectively. Including the sum $\sum_{p\, \subseteq (v_{\mathcal{L}} \cup \mathcal{L})}$ from \eq{eq:A.b3} we obtain
\begin{widetext}
\begin{equation} \label{eq:A.b5}
\begin{split}
\Delta N_{AB}(v_{\mathcal{L}}) = \frac{1}{4} \sum_{n = 0}^{N-1} (-1)^{x_n+y_n} \sum_{p\, \subseteq (v_{\mathcal{L}} \cup \mathcal{L})} \, \bigl|\{\bs{r}_n\}\cap p \bigr| \, \Bigl[ \bigl| \left\{\bs{r}_{n+1} \right\} \cap p \bigr| \bigl| \left\{\bs{r}_{n-1} \right\} \cap p \bigr| - \left(1-\bigl| \left\{\bs{r}_{n+1}\right\} \cap p \bigr|\right) \left(1-\bigl| \left\{\bs{r}_{n-1}\right\} \cap p \bigr| \right)\Bigr].
\end{split}
\end{equation}
\end{widetext}
Two evaluate the sum over $p$ in \eq{eq:A.b5}, we look at the two terms in the square brackets seperately. For the first term, we have
\begin{equation} \label{eq:A.b6}
\begin{split}
\sum_{p\, \subseteq V_{\mathcal{L}}} \bigl|\{\bs{r}_n\}\cap p \bigr|& \, \bigl| \left\{\bs{r}_{n+1} \right\} \cap p \bigr| \bigl| \left\{\bs{r}_{n-1} \right\} \cap p \bigr| = \\
=& \; \delta\bigl(\bs{\ell}_{o}(\bs{r}_n) \wedge \bs{\ell}_{i}(\bs{r}_n)-1\bigr).
\end{split}
\end{equation}
This relation can be understood in the following way: since all three $\bs{r}_{n-1},\bs{r}_{n},\bs{r}_{n+1}$ are supposed to be part of one plaquette $p$, the loop $\mathcal{L}$ needs to have a corner at $\bs{r}_n$, see \figc{fig:A.b1}{c}. Thus, $\bs{\ell}_{o}(\bs{r}_n) \wedge \bs{\ell}_{i}(\bs{r}_n) \neq 0$ needs to be finite. Furthermore, if there indeed is a corner of $\mathcal{L}$ at $\bs{r}_n$, there exists exactly one plaquette $p \subset \mathbb{Z}^2$ such that $\{\bs{r}_{n-1},\bs{r}_{n},\bs{r}_{n+1}\} \in p$. However, this plaquette will only be contained in $v_{\mathcal{L}} \cup \mathcal{L}$, and thus in the sum over $p$ in \eq{eq:A.b6}, if $\bs{\ell}_{o}(\bs{r}_n) \wedge \bs{\ell}_{i}(\bs{r}_n) = 1$, giving rise to the delta function. Note that changing the direction of $\mathcal{L}$ (via the inversion operator $\hat{I}_z$) nominally exchanges in- and outside of $\mathcal{L}$, see \figc{fig:A.b1}{b}.

Analogously, we obtain for the second term of \eq{eq:A.b5}
\begin{equation} \label{eq:A.b7}
\begin{split}
\sum_{p\, \subseteq V_{\mathcal{L}}} \bigl|\{\bs{r}_n\}\cap p \bigr|& \, \left(1-\bigl| \left\{\bs{r}_{n+1}\right\} \cap p \bigr|\right) \left(1-\bigl| \left\{\bs{r}_{n-1}\right\} \cap p \bigr| \right) = \\
= & \; \delta\bigl(\bs{\ell}_{o}(\bs{r}_n) \wedge \bs{\ell}_{i}(\bs{r}_n)+1\bigr).
\end{split}
\end{equation}
Inserting both \eq{eq:A.b6} and \eq{eq:A.b7} into \eq{eq:A.b5}, and using that 
\begin{equation} \label{eq:A.b8}
\begin{split}
\delta\bigl(\bs{\ell}_{o}(\bs{r}_n) \wedge \bs{\ell}_{i}(\bs{r}_n)-1\bigr) &- \delta\bigl(\bs{\ell}_{o}(\bs{r}_n) \wedge \bs{\ell}_{i}(\bs{r}_n)+1\bigr) = \\
&= \; \bs{\ell}_{o}(\bs{r}_n) \wedge \bs{\ell}_{i}(\bs{r}_n)
\end{split}
\end{equation}
due to $\bs{\ell}_{o}(\bs{r}_n) \wedge \bs{\ell}_{i}(\bs{r}_n) \; \in \; \{-1,0,1\}$, results in \eq{eq:A.b1} and thus completes our proof.

\section{Proof of \eq{eq:S3.5}} \label{sec:app4}

As stated in the main text, independent of the chosen boundary conditions, the following quantity is invariant under the dynamics of $\hat{H}_J$:
\begin{equation} \label{eq:A.c1}
\hat{\mathcal{Q}} = \sum_{\bs{r}} (-1)^{r_y+r_y} \bigl(  \hat{\bs{\ell}}_o(\bs{r})\wedge \hat{\bs{\ell}}_i(\bs{r})  \bigr),
\end{equation}
where
\begin{equation} \label{eqe:A.c2}
\begin{split}
\hat{\bs{\ell}}_o(\bs{r}) &= \sum_{\alpha \in \{\pm x, \pm y\}}\bs{e}_\alpha \; \hat{n}^{(l)}_{\bs{r},\alpha} \\
\hat{\bs{\ell}}_i(\bs{r}) &= \sum_{\alpha \in \{\pm x, \pm y\}}\bs{e}_\alpha \; \hat{n}^{(l)}_{\bs{r}-\bs{e}_\alpha,\alpha}.
\end{split}
\end{equation}

From the form $\hat{H}_J = \hat{H}^{(h)}_J + \hat{H}^{(l)}_J$ given in \eq{eq:14} and \eq{eq:15}, we notice that any local term in $\hat{H}^{(h)}_J$ creates or annihilates a trivial loop of length two that contains no corners. It is therfore directly verified that $[\hat{H}^{(h)}_J,\hat{\mathcal{Q}}]=0$.

To show that the remaining $\hat{H}^{(l)}_J$ also commutes with $\hat{\mathcal{Q}}$,
let us consider a local plaquette move $\hat{h}^{(l)}_p$ from $\hat{H}^{(l)}_J$ and show that $[\hat{h}^{(l)}_p,\hat{\mathcal{Q}}]=0$. Here, $p=\{\bs{r}_{p,1},\bs{r}_{p,2},\bs{r}_{p,3},\bs{r}_{p,4}\}$ labels the four sites of a given plaquette $p$ in counter-clockwise order (starting at the bottom left site) as defined in \eq{eq:A.b2}. According to \eq{eq:14}, the local term $\hat{h}^{(l)}_p$ is given either by 
\begin{equation} \label{eq:A.c44}
\hat{h}^{(l)}_p = \ket{\loopP} \bra{\loopPP} + h.c.,
\end{equation}
or
\begin{equation} \label{eq:A.c444}
\hat{h}^{(l)}_p = \ket{\loopPPP} \bra{\loopPPPP} + h.c.,
\end{equation}
and the following arguments proceed analogously for either choice.
Using
\begin{equation} \label{eq:A.c4}
[\hat{h}^{(l)}_p,\hat{\bs{\ell}}_o(\bs{r})\wedge \hat{\bs{\ell}}_i(\bs{r})] = [\hat{h}^{(l)}_p,\hat{\bs{\ell}}_o(\bs{r})] \wedge \hat{\bs{\ell}}_i(\bs{r}) + \hat{\bs{\ell}}_o(\bs{r})\wedge [\hat{h}^{(l)}_p,\hat{\bs{\ell}}_i(\bs{r})],
\end{equation}
we can compute
\begin{widetext}
\begin{equation} \label{eq:A.c5}
\begin{split}
& [\hat{h}^{(l)}_p,\hat{\mathcal{Q}}] = \sum_{\bs{r} \, \in \, p} (-1)^{r_x+r_y} [\hat{h}^{(l)}_p,\hat{\bs{\ell}}_o(\bs{r})\wedge \hat{\bs{\ell}}_i(\bs{r})] = \\
& = \Bigl( \hat{\bs{\ell}}_o(\bs{r}_{p,1})\wedge [\hat{h}^{(l)}_p,\hat{\bs{\ell}}_i(\bs{r}_{p,1})] + \hat{\bs{\ell}}_o(\bs{r}_{p,3})\wedge [\hat{h}^{(l)}_p,\hat{\bs{\ell}}_i(\bs{r}_{p,3})] - [\hat{h}^{(l)}_p,\hat{\bs{\ell}}_o(\bs{r}_{p,2})] \wedge \hat{\bs{\ell}}_i(\bs{r}_{p,2}) - [\hat{h}^{(l)}_p,\hat{\bs{\ell}}_o(\bs{r}_{p,4})] \wedge \hat{\bs{\ell}}_i(\bs{r}_{p,4}) \Bigr) +  \\
& \quad +  \Bigl( [\hat{h}^{(l)}_p,\hat{\bs{\ell}}_o(\bs{r}_{p,1})] \wedge \hat{\bs{\ell}}_i(\bs{r}_{p,1}) + [\hat{h}^{(l)}_p,\hat{\bs{\ell}}_o(\bs{r}_{p,3})] \wedge \hat{\bs{\ell}}_i(\bs{r}_{p,3}) - \hat{\bs{\ell}}_o(\bs{r}_{p,2})\wedge [\hat{h}^{(l)}_p,\hat{\bs{\ell}}_i(\bs{r}_{p,2})] - \hat{\bs{\ell}}_o(\bs{r}_{p,4})\wedge [\hat{h}^{(l)}_p,\hat{\bs{\ell}}_i(\bs{r}_{p,4})] \Bigr),
\end{split}
\end{equation}
\end{widetext}
where we have grouped the arising terms into two contributions marked by round brackets, which we are going to consider separately. 

For the term in the first round bracket of \eq{eq:A.c5}, the hard-core constraint implies, through direct evaluation,
\begin{widetext}
\begin{equation} \label{eq:A.c6}
\begin{split}
[\hat{h}^{(l)}_p,\hat{\bs{\ell}}_i(\bs{r}_{p,1})]\, =\, & \hat{h}^{(l)}_p \, \bigl(\bs{e}_y-\bs{e}_x\bigr) \times \delta\bigl(\hat{\bs{\ell}}_i(\bs{r}_{p,1})+\bs{e}_x\bigr)\, \delta\bigl(\hat{\bs{\ell}}_o(\bs{r}_{p,2})+\bs{e}_x\bigr)\, \delta\bigl(\hat{\bs{\ell}}_i(\bs{r}_{p,3})-\bs{e}_x\bigr)\, \delta\bigl(\hat{\bs{\ell}}_o(\bs{r}_{p,4})-\bs{e}_x\bigr) \, +\\
&+ \, \hat{h}^{(l)}_p \, \bigl(\bs{e}_x-\bs{e}_y\bigr) \times \delta\bigl(\hat{\bs{\ell}}_i(\bs{r}_{p,1})+\bs{e}_y\bigr)\, \delta\bigl(\hat{\bs{\ell}}_o(\bs{r}_{p,2})-\bs{e}_y\bigr)\, \delta\bigl(\hat{\bs{\ell}}_i(\bs{r}_{p,3})-\bs{e}_y\bigr)\, \delta\bigl(\hat{\bs{\ell}}_o(\bs{r}_{p,4})+\bs{e}_y\bigr) \\ \\
[\hat{h}^{(l)}_p,\hat{\bs{\ell}}_i(\bs{r}_{p,3})]\, =\, & ... = - [\hat{h}^{(l)}_p,\hat{\bs{\ell}}_i(\bs{r}_{p,1})] \\ \\
[\hat{h}^{(l)}_p,\hat{\bs{\ell}}_o(\bs{r}_{p,2})]\, =\, & \hat{h}^{(l)}_p \, \bigl(-\bs{e}_x-\bs{e}_y\bigr) \times \delta\bigl(\hat{\bs{\ell}}_i(\bs{r}_{p,1})+\bs{e}_x\bigr)\, \delta\bigl(\hat{\bs{\ell}}_o(\bs{r}_{p,2})+\bs{e}_x\bigr)\, \delta\bigl(\hat{\bs{\ell}}_i(\bs{r}_{p,3})-\bs{e}_x\bigr)\, \delta\bigl(\hat{\bs{\ell}}_o(\bs{r}_{p,4})-\bs{e}_x\bigr) \, +\\
&+ \, \hat{h}^{(l)}_p \, \bigl(\bs{e}_x+\bs{e}_y\bigr) \times \delta\bigl(\hat{\bs{\ell}}_i(\bs{r}_{p,1})+\bs{e}_y\bigr)\, \delta\bigl(\hat{\bs{\ell}}_o(\bs{r}_{p,2})-\bs{e}_y\bigr)\, \delta\bigl(\hat{\bs{\ell}}_i(\bs{r}_{p,3})-\bs{e}_y\bigr)\, \delta\bigl(\hat{\bs{\ell}}_o(\bs{r}_{p,4})+\bs{e}_y\bigr) \\ \\
[\hat{h}^{(l)}_p,\hat{\bs{\ell}}_o(\bs{r}_{p,4})]\, =\, & ... = - [\hat{h}^{(l)}_p,\hat{\bs{\ell}}_o(\bs{r}_{p,2})],
\end{split}
\end{equation}
\end{widetext}
see \figc{fig:1}{c} for an intuition about the terms appearing in \eq{eq:A.c6}. Inserting \eq{eq:A.c6} into the first term of \eq{eq:A.c5} yields
\begin{widetext}
\begin{equation} \label{eq:A.c7}
\begin{split}
\Bigl( \hat{\bs{\ell}}_o(\bs{r}_{p,1})\wedge [\hat{h}^{(l)}_p,\hat{\bs{\ell}}_i(\bs{r}_{p,1})] + \hat{\bs{\ell}}_o(\bs{r}_{p,3}) & \wedge [\hat{h}^{(l)}_p,\hat{\bs{\ell}}_i(\bs{r}_{p,3})] - [\hat{h}^{(l)}_p,\hat{\bs{\ell}}_o(\bs{r}_{p,2})] \wedge \hat{\bs{\ell}}_i(\bs{r}_{p,2}) - [\hat{h}^{(l)}_p,\hat{\bs{\ell}}_o(\bs{r}_{p,4})] \wedge \hat{\bs{\ell}}_i(\bs{r}_{p,4}) \Bigr) = \\
= \; \hat{h}^{(l)}_p\,  \Bigl( \bigl( \hat{\bs{\ell}}_o(\bs{r}_{p,1}) &- \hat{\bs{\ell}}_o(\bs{r}_{p,3}) \bigr) \wedge (\bs{e}_y-\bs{e}_x) + \bigl( \hat{\bs{\ell}}_i(\bs{r}_{p,4}) - \hat{\bs{\ell}}_i(\bs{r}_{p,2}) \bigr) \wedge (\bs{e}_x+\bs{e}_y) \Bigr) \times \\
& \times \, \Bigr[ \delta\bigl(\hat{\bs{\ell}}_i(\bs{r}_{p,1})+\bs{e}_x\bigr)\, \delta\bigl(\hat{\bs{\ell}}_o(\bs{r}_{p,2})+\bs{e}_x\bigr)\, \delta\bigl(\hat{\bs{\ell}}_i(\bs{r}_{p,3})-\bs{e}_x\bigr)\, \delta\bigl(\hat{\bs{\ell}}_o(\bs{r}_{p,4})-\bs{e}_x\bigr) \, - \\ 
& \qquad \qquad - \, \delta\bigl(\hat{\bs{\ell}}_i(\bs{r}_{p,1})+\bs{e}_y\bigr)\, \delta\bigl(\hat{\bs{\ell}}_o(\bs{r}_{p,2})-\bs{e}_y\bigr)\, \delta\bigl(\hat{\bs{\ell}}_i(\bs{r}_{p,3})-\bs{e}_y\bigr)\, \delta\bigl(\hat{\bs{\ell}}_o(\bs{r}_{p,4})+\bs{e}_y\bigr) \Bigr].
\end{split}
\end{equation}
\end{widetext}
While arranging the terms in \eq{eq:A.c7} we have used the fact that
\begin{equation} \label{eq:A.c8}
[\hat{h}^{(l)}_p,\hat{\bs{\ell}}_{i/o}(\bs{r})] \neq 0 \Rightarrow [\hat{h}^{(l)}_p,\hat{\bs{\ell}}_{o/i}(\bs{r})] = 0.
\end{equation}
Let us now focus on the round bracket of the right hand side of \eq{eq:A.c7}: There are in total $34$ different possible (i.e. compatible with the hard core constraint) combinations of eigenvalues of the four operators $\hat{\bs{\ell}}_o(\bs{r}_{p,1}), \hat{\bs{\ell}}_o(\bs{r}_{p,3}), \hat{\bs{\ell}}_i(\bs{r}_{p,2}), \hat{\bs{\ell}}_i(\bs{r}_{p,4})$  appearing in \eq{eq:A.c7}:
\begin{itemize}
\item 16 possibilities from $\hat{\bs{\ell}}_o(\bs{r}_{p,1}) \in \{-\bs{e}_x,-\bs{e}_y\}$, $\hat{\bs{\ell}}_o(\bs{r}_{p,3}) \in \{\bs{e}_x,\bs{e}_y\}$, $\hat{\bs{\ell}}_i(\bs{r}_{p,2}) \in \{-\bs{e}_x,\bs{e}_y\}$, $\hat{\bs{\ell}}_i(\bs{r}_{p,4}) \in \{\bs{e}_x,-\bs{e}_y\}$ independently.
\item 4 possibilities from $\hat{\bs{\ell}}_o(\bs{r}_{p,1}) = \bs{e}_x$, $\hat{\bs{\ell}}_i(\bs{r}_{p,2}) = \bs{e}_x$, $\hat{\bs{\ell}}_o(\bs{r}_{p,3}) \in \{\bs{e}_x,\bs{e}_y\}$, $\hat{\bs{\ell}}_i(\bs{r}_{p,4}) \in \{\bs{e}_x,-\bs{e}_y\}$.
\item 4 possibilities from $\hat{\bs{\ell}}_o(\bs{r}_{p,1}) = \bs{e}_y$, $\hat{\bs{\ell}}_i(\bs{r}_{p,4}) = \bs{e}_y$, $\hat{\bs{\ell}}_o(\bs{r}_{p,3}) \in \{\bs{e}_x,\bs{e}_y\}$, $\hat{\bs{\ell}}_i(\bs{r}_{p,2}) \in \{-\bs{e}_x,\bs{e}_y\}$.
\item 4 possibilities from $\hat{\bs{\ell}}_o(\bs{r}_{p,3}) = -\bs{e}_x$, $\hat{\bs{\ell}}_i(\bs{r}_{p,4}) = -\bs{e}_x$, $\hat{\bs{\ell}}_o(\bs{r}_{p,1}) \in \{-\bs{e}_x,-\bs{e}_y\}$, $\hat{\bs{\ell}}_i(\bs{r}_{p,2}) \in \{-\bs{e}_x,\bs{e}_y\}$.
\item 4 possibilities from $\hat{\bs{\ell}}_o(\bs{r}_{p,3}) = -\bs{e}_y$, $\hat{\bs{\ell}}_i(\bs{r}_{p,2}) = -\bs{e}_y$, $\hat{\bs{\ell}}_o(\bs{r}_{p,1}) \in \{-\bs{e}_x,-\bs{e}_y\}$, $\hat{\bs{\ell}}_i(\bs{r}_{p,4}) \in \{\bs{e}_x,-\bs{e}_y\}$.
\item 1 possibility from $\hat{\bs{\ell}}_o(\bs{r}_{p,1}) = \bs{e}_x$, $\hat{\bs{\ell}}_i(\bs{r}_{p,2}) = \bs{e}_x$, $\hat{\bs{\ell}}_o(\bs{r}_{p,3}) = -\bs{e}_x$, $\hat{\bs{\ell}}_i(\bs{r}_{p,4}) = -\bs{e}_x$.
\item 1 possibility from $\hat{\bs{\ell}}_o(\bs{r}_{p,1}) = \bs{e}_y$, $\hat{\bs{\ell}}_i(\bs{r}_{p,4}) = \bs{e}_y$, $\hat{\bs{\ell}}_o(\bs{r}_{p,3}) = -\bs{e}_y$, $\hat{\bs{\ell}}_i(\bs{r}_{p,2}) = -\bs{e}_y$.
\end{itemize}
It is then a straightforward task to go through all $34$ listed possibilities and check that in each case, the round bracket on the right hand side of \eq{eq:A.c7}, and thus the left hand side of \eq{eq:A.c7} itself, vanishes.

We can then repeat this derivation analogously for the second round bracket on the right hand side of \eq{eq:A.c5}, which consequently also vanishes, such that overall, we indeed find $[\hat{h}_p,\hat{\mathcal{Q}}]=0$, proving \eq{eq:A.c1}.

\section{Conservation of chiral subcharges} \label{sec:app1}

In this Appendix we formally show the invariance of the chiral subcharges $\hat{Q}_\phi$ under the Hamiltonian $\hat{H}_J=\hat{H}^{(l)}_J+\hat{H}^{(h)}_J$ given in \eq{eq:14} and \eq{eq:15}, as argued for in Sec.~\ref{sec:2}. 

Taking care of $\hat{H}^{(l)}_J$ first, we verify through direct inspection of the possible loop moves in \eq{eq:14} that $[\hat{H}^{(l)}_J,\hat{q}_{\bs{r}}(\phi)]=0$ for all $\bs{r}$ and $\phi$, with $\hat{q}_{\bs{r}}(\phi)$ from \eq{eq:19}. This immediately implies $[\hat{H}^{(l)}_J,\hat{Q}^{}_{\phi}]=0$ for all $\phi$.

Moving on to $\hat{H}^{(h)}_J$ from \eq{eq:15} next, we use that $[\hat{l}^\dagger_{\bs{r},\alpha}\, \hat{l}^\dagger_{\bs{r}+\bs{e}_\alpha,-\alpha},\hat{\phi}_{\bs{r}^\prime}]=0$ and thus also $[\hat{l}^\dagger_{\bs{r},\alpha}\, \hat{l}^\dagger_{\bs{r}+\bs{e}_\alpha,-\alpha},\,\hat{q}_{\bs{r}^\prime}(\phi)]=0$ for all $\bs{r},\bs{r}^\prime$. We can then compute the commutator
\begin{widetext}
\begin{equation} \label{eq:22}
\begin{split}
[\hat{H}^{(h)}_J,\hat{Q}_\phi] = & \sum_{\bs{r},\alpha} (-1)^{r_x+r_y} \Biggl\{ \hat{l}^\dagger_{\bs{r},\alpha}\,\hat{l}^\dagger_{\bs{r}+\bs{e}_\alpha,-\alpha} \, \hat{h}_{\bs{r}}\,\hat{h}_{\bs{r}+\bs{e}_\alpha}\; \Bigl[\delta\left(\hat{\phi}_{\bs{r}+\bs{e}_\alpha}-\phi\right) - \delta\left(\hat{\phi}_{\bs{r}}-\phi\right) \Bigr]\Biggr\} + \\
+\, & \sum_{\bs{r},\alpha} (-1)^{r_x+r_y} \Biggl\{ \hat{l}_{\bs{r},\alpha}\,\hat{l}_{\bs{r}+\bs{e}_\alpha,-\alpha} \, \hat{h}^\dagger_{\bs{r}}\,\hat{h}^\dagger_{\bs{r}+\bs{e}_\alpha}\; \Bigl[ \delta\left(\hat{\phi}_{\bs{r}}-\phi\right) - \delta\left(\hat{\phi}_{\bs{r}+\bs{e}_\alpha}-\phi\right) \Bigr]\Biggr\}.
\end{split}
\end{equation}
\end{widetext}
To demonstrate that \eq{eq:22} indeed vanishes, we have to show that 
\begin{equation} \label{eq:A.a1}
\hat{h}_{\bs{r}}\,\hat{h}_{\bs{r}+\bs{e}_\alpha}\,\hat{\phi}_{\bs{r}} = \hat{h}_{\bs{r}}\,\hat{h}_{\bs{r}+\bs{e}_\alpha}\,\hat{\phi}_{\bs{r}+\bs{e}_\alpha}
\end{equation} 
for both $\alpha =x,y$, which directly yields zero upon insertion into \eq{eq:22}. Setting $\alpha=x$, we see from the definition of $\hat{\phi}_{\bs{r}}$ in \eq{eq:18} that 
\begin{equation} \label{eq:A.a2}
\begin{split}
\hat{h}_{\bs{r}}\,\hat{h}_{\bs{r}+\bs{e}_x}\,\hat{\phi}_{\bs{r}} &- \hat{h}_{\bs{r}}\,\hat{h}_{\bs{r}+\bs{e}_x}\,\hat{\phi}_{\bs{r}+\bs{e}_x} = \\
=\, &\hat{h}_{\bs{r}}\,\hat{h}_{\bs{r}+\bs{e}_x} \bigl( \hat{n}^{(l)}_{\bs{r},y}-\hat{n}^{(l)}_{\bs{r}+\hat{e}_y,-y} \bigr) = 0.
\end{split}
\end{equation} 
The last equality in \eq{eq:A.a2} is due to the hard core constraint: if there is a charge at site $\bs{r}$, then there cannot be a loop segment running through $\bs{r}$. This proves \eq{eq:A.a1} for $\alpha = x$. 
For $\alpha=y$, \eq{eq:A.a1} can be verified in the following way: Assume $\hat{n}^{(h)}_{\bs{r}+\hat{e}_y} = 1$, i.e. an interlayer charge occupies the site at $\bs{r}+\bs{e}_y$ (otherwise, \eq{eq:A.a1} yields zero immediately). Consider a loop segment that gives a contribution $+1$ to $\hat{\phi}_{\bs{r}}$. This segment enters the horizontal string to the left of $\bs{r}+\bs{e}_y$ from below and has two options: 1) It leaves the string going upwards, therefore also giving a contribution $+1$ to $\hat{\phi}_{\bs{r}+\bs{e}_y}$. 2) It leaves the string going downwards, thus giving no contribution to $\hat{\phi}_{\bs{r}+\bs{e}_y}$, but yielding an additional contribution $-1$ to $\hat{\phi}_{\bs{r}}$, and therefore net contribution zero. In both cases, $\hat{\phi}_{\bs{r}} = \hat{\phi}_{\bs{r}+\bs{e}_y}$. The same argument holds for loop segments running in the opposite direction. Note that this argument relies on $\hat{n}^{(h)}_{\bs{r}+\bs{e}_y} = 1$, otherwise a loop might enter the horizontal string without leaving it, by running directly through site $\bs{r}+\bs{e}_y$. This proves \eq{eq:A.a1} for $\alpha = y$.

Intuitively, the proof can be summarized as follows: Loop-dynamics can deform the shape, position, and number of loops in the system, but never change the net charge contained inside the interior of the loops. The dynamics of charges occurs as creation and annihilation of oppositely charged interlayer dimers on neighboring lattice sites, which thus are both enclosed by the same net chirality.

\section{Proof of \eq{eq:21}} \label{sec:app2}
In order to prove \eq{eq:21} of the main text, we first introduce a formal definition of the `interior' $v_\mathcal{L}$ of a closed loop $\mathcal{L}$ on open boundaries, as illustrated in \fig{fig:3}.
To do so, let us first define a string operator
\begin{equation} \label{eq:Abb.1}
\hat{\phi}^{(\mathcal{L})}_{\bs{r}} = \sum_{\substack{r^\prime_x = 0 \\ (r^\prime_x,r_y) \in \mathcal{L}}}^{r_x-1} \Bigl[ \hat{n}^{(l)}_{(r^\prime_x,r_y),y}
- \hat{n}^{(l)}_{(r^\prime_x,r_y+1),-y} \Bigr],
\end{equation}
which is similar to \eq{eq:18}, but sums only over sites contained in the set $\mathcal{L}=\{\bs{r}_0,...,\bs{r}_{|\mathcal{L}|-1}\}$. We can then define $\tilde{v}_\mathcal{L}=\{\bs{r} \notin \mathcal{L}:\; |\phi^{(\mathcal{L})}_{\bs{r}}|=1\}$, which gives us a set of sites enclosed by $\mathcal{L}$ (but excluding $\mathcal{L}$ itself) on open boundary conditions, see \fig{fig:3}. Notice however that due to the chirality of $\mathcal{L}$, its interior should become the complement of $\tilde{v}_\mathcal{L}$ upon reversing the chirality (again excluding $\mathcal{L}$ itself). On a $V = [0,L_x] \times [0,L_y]$ lattice, the interior of $\mathcal{L}$ is thus given by
\begin{equation} \label{eq:Abb.2}
v_{\mathcal{L}} = \frac{1}{2}\Bigl(1+\phi^{(\mathcal{L})}_{\bs{r} \in \tilde{v}_\mathcal{L}}\Bigr)\, \tilde{v}_\mathcal{L} \cup \frac{1}{2}\Bigl(1-\phi^{(\mathcal{L})}_{\bs{r} \in \tilde{v}_\mathcal{L}}\Bigr)\, V \backslash (\tilde{v}_\mathcal{L} \cup \mathcal{L}).
\end{equation}
Note that $\phi^{(\mathcal{L})}_{\bs{r} \in \tilde{v}_\mathcal{L}} \in \{-1,+1\}$ is independent of the chosen $\bs{r} \in \tilde{v}_\mathcal{L}$.\\

With these definitions, we can start from the right hand side of \eq{eq:21} and rewrite
\begin{equation} \label{eq:Abb.3}
\begin{split}
& \sum_\phi \; \phi \; \hat{Q}_\phi = \sum_{\bs{r}}(-1)^{r_x+r_y} \sum_{\phi} \phi\, \hat{q}_{\bs{r}}(\phi) = \\
& \qquad = \sum_{\bs{r}}(-1)^{r_x+r_y} \sum_{\mathcal{L}} \phi^{(\mathcal{L})}_{\bs{r}^\prime \in \tilde{v}_{\mathcal{L}}} \; \bigl|\{\bs{r}\} \cap \tilde{v}_{\mathcal{L}}\bigr| \; \hat{n}^{(h)}_{\bs{r}},
\end{split}
\end{equation}
where $\bigl|\{\bs{r}\} \cap \tilde{v}_{\mathcal{L}}\bigr| \in \{0,1\}$ measures whether site $\bs{r}$ is contained within the set $\tilde{v}_{\mathcal{L}}$ or not. Again, $\phi^{(\mathcal{L})}_{\bs{r}^\prime \in \tilde{v}_\mathcal{L}} \in \{-1,+1\}$ is independent of $\bs{r}^\prime$. The sum $\sum_{\mathcal{L}}$ extends over all loops $\mathcal{L}$ within the transition graph of a given dimer configuration. Rearranging the sums in \eq{eq:Abb.3} we obtain
\begin{equation} \label{eq:Abb.4}
\begin{split}
\sum_\phi \; \phi \; \hat{Q}_\phi &= \sum_{\mathcal{L}} \phi^{(\mathcal{L})}_{\bs{r}^\prime \in \tilde{v}_{\mathcal{L}}} \sum_{\bs{r}} (-1)^{r_x+r_y} \; \bigl|\{\bs{r}\} \cap \tilde{v}_{\mathcal{L}}\bigr| \; \hat{n}^{(h)}_{\bs{r}} = \\
&= \sum_{\mathcal{L}} \phi^{(\mathcal{L})}_{\bs{r}^\prime \in \tilde{v}_{\mathcal{L}}} \bigl[ N_A(\tilde{v}_{\mathcal{L}}) - N_B(\tilde{v}_{\mathcal{L}}) \bigr] = \\
&= \sum_{\mathcal{L}} \bigl[ N_A(v_{\mathcal{L}}) - N_B(v_\mathcal{L}) \bigr] = \hat{\mathcal{Q}},
\end{split}
\end{equation}
where $N_{A/B}(s)$ denotes the number of $A/B$ sublattice sites contained within a set $s \subset \mathbb{Z}^2$. From the first to the second line, we have used that an imbalance in the number of positive/negative charges $\hat{n}^{(h)}_{\bs{r}}$ on the sites within $\tilde{v}_{\mathcal{L}}$ is directly reflected in the imbalance of the number of $A/B$ sublattice sites within $\tilde{v}_{\mathcal{L}}$. This is due to the fact that all loops $\mathcal{L}^\prime$, which may potentially be contained within $\tilde{v}_{\mathcal{L}}$ for a given transition graph, are of even length and thus contain the same amount of $A/B$ sublattice sites. From the second to the third line in \eq{eq:Abb.4}, we have used that the system has even lengths $L_x,L_y$ in both directions, which implies $N_A(V)-N_B(V)=0$ for the entire lattice $V$ under considerations. This completes our proof of \eq{eq:21}.

\section{Symmetry of $\hat{H}_J$} \label{sec:app5}

\begin{center}
\textit{Symmetry of $\hat{H}_J$}
\end{center}

The Hamiltonian $\hat{H}_J = -J \sum_p \hat{h}_p$ can be demonstrated to have a symmetric spectrum: We define an operator
\begin{equation} \label{eq:spec1}
\hat{P} =  (-1)^{\sum_{\{\bs{r}|\, r_x+r_y = 0\, \mathrm{mod}\, 2\}} \bigl( \hat{n}^{(d)}_{\bs{r},x}+\hat{n}^{(d)}_{\bs{r},z} \bigr) },
\end{equation}
which yieds the parity of the total number of dimers emerging into either $x$- or $z$-direction from lattice sites $\bs{r}=(r_x,r_y,r_z)$ that fulfill $r_x+r_y = 0 \; \mathrm{mod}\, 2$. It can straightforwardly be verified that each plaquette of the lattice contains either one or three, i.e. an \textit{odd} number of bonds that contribute to the parity of \eq{eq:spec1}. Therefore, $\{\hat{P},\hat{h}_p\}=0$ for all plaquettes $p$ and thus the operator $\hat{P}$ anticommutes with the Hamiltonian, $\{\hat{P},\hat{H}_J\}=0$. Hence, for every eigenstate $\hat{H}_J \ket{\psi} = E_\psi \ket{\psi}$ there exists a corresponding state $\ket{\psi^\prime} = \hat{P}\ket{\psi}$ with opposite energy $\hat{H}_J \ket{\psi^\prime} = -E_\psi \ket{\psi^\prime}$. Note that this argument is independent of the spatial dimension of the dimer model.

As a consequence of the symmetric spectrum, each product state in the basis of dimer occupation numbers has energy expectation value zero, and thus formally corresponds to `infinite temperature'.

\bibliography{hopfions}

\end{document}